\documentclass[12pt]{article}
\newmuskip\pFqmuskip
\newcommand*\pFq[6][8]{%
	\begingroup 
	\pFqmuskip=#1mu\relax
	\mathcode`\,=\string"8000
	\begingroup\lccode`\~=`\,
	\lowercase{\endgroup\let~}\pFqcomma
	{}_{#2}F_{#3}{\left(\left.\genfrac..{0pt}{}{#4}{#5}\right|#6\right)}%
	\endgroup
}
\newcommand{\pFqcomma}{\mskip\pFqmuskip}

\pdfoutput=1
\usepackage{graphicx}
\usepackage{jheppub}
\usepackage{tikz}
\usepackage{hyperref}
\usepackage{amsmath}

\usepackage[utf8]{inputenc}
\usetikzlibrary{patterns.meta}

\usetikzlibrary{patterns}

\newenvironment{psmallmatrix}
{\left[\begin{smallmatrix}}
	{\end{smallmatrix}\right]}

\title{A note on RG domain wall   between successive   $A_2^{(p)}$ minimal models}

\usepackage{mathtools}

\DeclarePairedDelimiterX\braket[2]{\langle}{\rangle}{#1 \delimsize\vert #2}

\author[]{Armen Poghosyan\quad}
\author[]{Hasmik Poghosyan }

\affiliation[]{Yerevan Physics Institute\\
Alikhanian Br. 2, 0036 Yerevan, Armenia}

\emailAdd{armenpoghos@yerphi.am\,\,}
\emailAdd{h.poghosyan@yerphi.am}

\abstract{ We investigate the RG domain wall between   neighboring $A_2^{(p)}$ minimal   CFT models and
 establish the map between UV and IR fields  (matrix of mixing coefficients). A particular  RG invariant set of 
 six primary and four descendant  fields is analyzed   in full details. Using the algebraic construction
 of the RG domain wall  we compute  the UV/IR mixing matrix.  
  To test our results we show  that  it
diagonalizes the   matrix of anomalous dimensions previously known from perturbative analysis.
It is important to note that the diagonalizing matrix can not be found from 
perturbative analysis solely due to degeneracy of anomalous dimensions.
The same mixing coefficients are used  to explore anomalous W-weights as well.
}

\begin{document}
\maketitle
\newcommand{\ie}{{\it i.e.\ }}
\def\bea{\begin{eqnarray}}
\def\eea{\end{eqnarray}}

\section*{Introduction}
In  \cite{Zamolodchikov:1987ti}  using  perturbation analysis  A.~Zamolodchikov
investigated the RG flow between  minimal models
${\cal M}_p$  and  ${\cal M}_{p-1}$  initiated by the
relevant field $\phi_{1,3}$. The next  to leading  order computations are  performed in \cite{Poghossian:2013fda}. These results were extended
for  ${\cal N}=1$ super-minimal models \cite{Poghossian:1987ngr,Kastor:1988ef, Crnkovic:1989gy,Ahn:2014rua}. 

A much broader class of such RG flows is known  to exist \cite{Crnkovic:1989gy,Ravanini:1992fs} for some CFT coset models,  among them are $W$-algebra invariant   minimal models $A_{n-1}^{(p)}$   \cite{Zamolodchikov:1985wn,lukyanov1991additional}.  
It is argued in  \cite{lukyanov1991additional} that under certain relevant perturbation 
$A_{n}^{(p)}$ flows to $A_{n}^{(p-1)}$.
In particular   $A_{2}^{(p)}$ flows to $A_{2}^{(p-1)}$ when  perturbed by the relevant field  $\phi \begin{psmallmatrix}	1 & 2\\	1 &2	\end{psmallmatrix}$.
In the  recent works  \cite{Poghosyan:2022mfw,Poghosyan:2022ecv} 
the matrices of anomalous dimensions   for three RG invariant classes of local  fields were computed explicitly.  This allowed to derive the UV/IR  mixing coefficients for the first two   classes. The third class besides primaries includes   secondary fields some having  differing  holomorphic and  anti-holomorphic  parts.
The knowledge of  anomalous dimensions for this class   happens to be  not sufficient to fix
all mixing coefficients. The reason is the presence of  four-fold (approximate) degeneracy of anomalous dimensions.
 In \cite{Poghosyan:2022mfw} the mixing coefficients for the first two classes have been successfully checked against algebraic RG domain wall approach
 suggested in \cite{Gaiotto:2012np}. Similar computation  for the third RG invariant class of fields, which would   specify respective mixing coefficients completely,     was left  open  due to substantial computational difficulties. 
In this work we have established few simplifying rules applicable for calculation of matrix elements in coset theories which allowed us to derive UV/IR mixing coefficients using RG domain wall approach.

Thus our  result provides a strong additional tests for the RG domain wall conjecture \cite{Gaiotto:2012np}.
Remind that the RG domain wall construction of \cite{Gaiotto:2012np} is supposed to be applicable 
for a rather large class of theories, with some cases being cheeked successfully  \cite{Poghossian:2013fda, Poghosyan:2013qta, Poghosyan:2014jia,Konechny:2014opa, Brunner:2015vva,Poghosyan:2022mfw}.
A similar construction is considered also for  boundary RG flows   \cite{Konechny:2016eek, Konechny:2020jym}.
In \cite{Brehm:2015lja} entanglement entropy is calculated in presence of the RG domain wall. Also 
such  domain walls are studied numerically  in  two and higher  dimensions in   \cite{Gliozzi:2015qsa}.

To conclude in  section  \ref{RGwall}  we have computed the pairings between UV/IR fields (see formulae  (\ref{uv10}) and (\ref{ir10}))  in presence of RG domain wall.
 For large level $k$ we have checked that  they constitute an orthogonal matrix (denoted by $R$ see section \ref{Mix_coef}), which
diagonalizes the matrix of anomalous dimensions  computed in \cite{Poghosyan:2022mfw}. In the same paper a definition
for    anomalous W-weights matrix (The W-analog for anomalous dimensions) was given.
Using RG domain wall approach we have computed this matrix in leading order of 
coupling constant and have shown that its matrix elements corresponding to
primary fields coincide with those obtained from perturbation theory.


\section{A concise review of $W_3$ minimal models and  RG flows}
\label{review}
In a CFT the energy-momentum tensor
has two nonzero components  namely the holomorphic component $T(z)$ and  its anti-holomorphic counterpart  $\bar{T}(\bar{z})$ with conformal dimensions $(2,0)$ and $(0,2)$ respectively.
We are interested in theories with additional  $W_3$ algebra symmetry generated by rank three symmetric tensor  currents with components   $W(z)$ and  $\bar{W}(\bar{z})$ of dimensions $(3,0)$ and $(0,3)$.
The OPE rules these fields obey read
{\footnotesize
	\begin{eqnarray}
    \nonumber
	&&	T(z)T(0)=\frac{c/2}{z^4}+\frac{2T(0)}{z^2}+\frac{T^\prime(0)}{z}  + \cdots , 
	\\ \label{TGOPE}
	&&     T(z)W(0)=\frac{3W(0)}{z^{2}}+\frac{W^\prime(0)}{z} +\cdots\,,\\
	\nonumber
	&&	W(z)W(0)= \frac{c/3}{z^6}+\frac{2T(0)}{z^4} + \frac{T^\prime(0)}{z^3}+\frac{1}{z^2}\left(\frac{15c+66}{10(22+5c)}T^{\prime\prime}(0)+\frac{32}{22+5c}\Lambda(0)\right)\\
	&& \qquad \qquad\qquad\qquad \qquad\qquad\qquad \qquad
	+\frac{1}{z}\left(\frac{1}{15}T^{\prime\prime\prime}(0)+\frac{16}{22+5c}\Lambda^{\prime}(0)\right) +\cdots  .\nonumber
	\end{eqnarray}
}
Here the  quasi-primary field $\Lambda(z)$ is  defined as
\bea
\nonumber
\Lambda(z)=:\!\!TT\!\!:(z)-\frac{3}{10}T^{\prime\prime}(z)\,.
\eea
These fields can be expanded in  Laurent series as
\begin{equation}
T(z)=\sum \limits_{n=-\infty}^{+\infty} \frac{L_n}{z^{n+2}}\,,
\quad
W(z)=\sum \limits_{n=-\infty}^{+\infty} \frac{W_n}{z^{n+3}}\,,
\quad
\Lambda(z)=\sum \limits_{n=-\infty}^{+\infty} \frac{\Lambda_n}{z^{n+4}}\,.
\end{equation}
The  Laurent coefficients are subject to the commutation
$W_3$
algebra relations  which can be derived from  OPE  rules (\ref{TGOPE})
{\footnotesize
	\begin{eqnarray}
	\nonumber
	&&[L_n,L_m] = (n-m)L_{n+m} +\frac{c}{12}(n^3-n)
	\delta_{n+m,0} \, , \\
	\label{comLW}
	&&\left[L_n, W_m\right] =\left(2n-m\right)W_{n+m} \, , \\
	\nonumber
	&&\left[W_n, W_m\right] =\alpha(n,m) L_{n+m}+\frac{16 (n-m)}{22+5c
	}\Lambda_{n+m}+\frac{c}{360}(n^2-4)(n^2-1)n\delta_{n+m,0},
	\, \,\,\,\,
	\end{eqnarray}
}

where 
\bea
&&\alpha(n,m)=(n-m)\left(\frac{1}{15}(n+m+2)(n+m+3)-\frac{1}{6}(n+2)(m+2)\right)\,,
\nonumber\\
&&\Lambda_{n}=d_nL_n+\sum_{-\infty}^{+\infty}:\!\! L_mL_{n-m}\!\!: \,.
\eea
Here $::$ means the normal ordering (i.e operators with smaller index come first) and
\bea
d_{2m}=\frac{1}{5}\left(1-m^2\right)\,,
\quad
d_{2m-1}=\frac{1}{5}\left(1+m\right)\left(2-m\right)\,.
\eea
In this paper we are interested in unitary  minimal models $A_2^{(p)}$ with Virasoro  central charge
\bea
\label{cp}
c_p=2-\frac{24}{p(p+1)}\,.
\eea
The  primary fields $\phi \begin{psmallmatrix}	n & m\\	n' & m'	 \end{psmallmatrix}$  of this theory are labeled by four integers 
$n$, $n'$, $m$, $m'$  subject to  constraints
\[n+n'\le p-1, \qquad  m+m'\le p\,.\]
Besides fields
\bea
\phi  \begin{psmallmatrix} n & m\\  n' & m'	\end{psmallmatrix} 
\quad {\rm and} \quad
\phi \begin{psmallmatrix} n' & m'\\p-n -n' & p+1-m-m'	\end{psmallmatrix}
\eea
should be  identified, since they have the same  Virasoro and $W$ dimensions (easily seen from explicit formulae below).

Primary fields have conformal dimensions
{\footnotesize
	\bea
	\label{dim}
	\Delta
	\begin{psmallmatrix}
		n & m\\
		n' & m'	
	\end{psmallmatrix}= 	
	\frac{((p+1) (n-n')-p (m-m'))^2+3 ((p+1) (n+n')-p (m+m'))^2-12}{12 p (p+1)}
	\,.
	\eea 
}
and corresponding $w$ weights 
{\footnotesize
	\bea 
	\label{w3weight}
	w\begin{psmallmatrix} n & m\\  n' & m'	\end{psmallmatrix}
	&=&\sqrt{\frac{2}{3}}((p+1) (n'-n)-p (m'-m))\times\\
	&\times&\frac{  ((p+1) (n+2 n')-p (m+2 m')) ((p+1) (2 n+n')-p (2 m+m'))}{9 p (p+1) \sqrt{(2 p+5) (2 p-3)}}\,.\nonumber
	\eea 
}
We will initiate  the RG flow with the primary  field
\bea
\label{pertField}
\phi^{(p)} \begin{psmallmatrix}	1 & 2\\	1 &2	\end{psmallmatrix}
 =\varphi (x)\,,
\eea 
which  is  relevant
\bea 
\label{int_field_dim}
\Delta \begin{psmallmatrix}	1 & 2\\	1 &2	\end{psmallmatrix}=
\frac{p-2}{p+1}\equiv 1-\epsilon<1 \,,
\quad
\epsilon=\frac{3}{p+1}
\eea
 and $w$-neutral
\bea 
\nonumber
w\begin{psmallmatrix}	1 & 2\\	1 &2	\end{psmallmatrix}=0\,.
\eea
Consider a subset of fields generated from our field   $\varphi$ 
with the help of OPE.  It is possible to show that in this subset 
 $\varphi$ will be the only relevant field besides the identity \footnote{In fact  $\varphi$ is the only field with such property.}.
  This  allows one   to construct a consistent perturbed CFT with a single coupling constant:
\bea
\label{action}
A=A_{CFT}+g \int \varphi (x)d^2x \,.
\eea 
As we see from (\ref{int_field_dim}) for large  $p>>1$, the perturbing field is only slightly 
relevant so that  the conformal perturbation theory becomes applicable along a large portion of the
RG flow. In  case of positive values of  the coupling $g >0$   it was shown in \cite{lukyanov1991additional} that   our initial theory $A_2^{(p)}$ flows to 
$A_2^{(p-1)}$. In \cite{Poghosyan:2022mfw} three  classes of RG invariant  field sets  were identified 
and their mixing coefficients for  two of these classes were  found. Investigation of the  remaining  largest class  (see (\ref{uv10})),  is
the focus of this paper.


\section{Current algebra}
\label{Coset_construction}
The WZNW models enjoy extended chiral   symmetry  generated by 
 Virasoro  and  current algebras. We are interested in $su(3)$ WZNW models. The colored part of currents we will denote as 
 $E^{ij}(z)$, $i \neq j$ and  $i,j\in 1,2,3$, while the  Cartan part as $H^i(z)$  with constraint 
 \bea
 H^1(z)+H^2(z)+H^3(z)=0\,.
 \eea
  Here are the OPE relations
 \bea
 \nonumber
 &&E^{ij}(z)E^{rl}(0)=\frac{k}{z^2}\delta^{il}\delta^{jr}+\frac{1}{z}\left(\delta^{jr}E^{il}(0)-\delta^{li}E^{rj}(0)\right)+\dots\,,
 \\
 \label{KM_OPE}
 &&E^{ij}(z)H^{r}(0)=\frac{1}{z}\left(\delta^{jr}E^{ir}(0)-\delta^{ri}E^{rj}(0)\right)+\dots\,,
 \\
 \nonumber
 &&H^{i}(z)H^{j}(0)=\frac{k}{z^2} \left(\delta_{ij}-\frac{1}{3}\right)+\dots\,.
 \eea
 Where the level $k=1,2,3....$ specifies a model.
 Since $E^{ij}$ are  primary fields with dimension one we also have
 \bea
 T(z)E^{ij}(0)=\frac{E^{ij}(0)}{z^2}+\frac{\partial E^{ij}(0)}{z}+\dots\,.
 \eea
The coefficients  of Laurent expansion 
\bea
E^{ij}(z)=\sum \limits_{n=-\infty}^{\infty} \frac{E^{ij}_n}{z^{n+1}}\,,
\quad
H^{l}(z)=\sum \limits_{n=-\infty}^{\infty} \frac{H^{l}_n}{z^{n+1}}
\eea
satisfy the ${\rm Ka\breve{c}-Moody}$ algebra
\bea
\nonumber
&&\left[E_n^{ij}E_m^{rl}\right]=\delta^{jr}E_{n+m}^{il}-\delta^{li}E_{n+m}^{rj}+kn\delta_{n+m}\delta^{il}\delta^{jr}\,,
\\
\label{KM_alg}
&&\left[E_n^{ij}H_m^{r}\right]=\delta^{jr}E_{n+m}^{ir}-\delta^{ri}E_{n+m}^{rj}\,,
\\
\nonumber
&&\left[H_n^{i}H_m^{j}\right]=k n \delta_{n+m}\left(\delta_{ij}-\frac{1}{3}\right)\,.
\eea
Notice  that
\[
\sum_{i=1}^3H^i_n=0\,.
\]
The Virasoro generators can be expressed in terms of  current algebra  trough the Sugawara construction
\begin{align}
\label{Ln_WZW_gen}
L_n=\frac{1}{2(k+3)}\sum_{l\in Z}\sum_{i,j=1}^3:E_l^{ij}E_{n-l}^{ji}:\,.
\end{align}
The normal ordering prescription  is the standard one: 
the generator with larger loop index should be moved to the right.

Using (\ref{KM_alg}) and  (\ref{Ln_WZW_gen}) one can get convinced that the 
 Virasoro central charge is
\begin{eqnarray}
c_k=\frac{8k}{k+3} \,.
\label{KM central charge}
\end{eqnarray}
The primary fields of the theory, denoted as $\phi_{\lambda}$,  are labeled by the
highest weights 
\bea
\label{lambda}
\lambda_{n,n'}=(n-1)\omega_1+(n'-1)\omega_2\,,
\eea
where $n$, $n'$ are positive integers.
We will represent 
roots, weights and Cartan elements of our algebra  as  3-component vectors
such that the sum of
components is zero.
 In this notation the highest weights
of  fundamental and anti-fundamental representations take the form  
\bea
\label{omega1_2}
\omega_1=\left(
\begin{array}{r}
	2/3\\
	-1/3\\
	-1/3	
\end{array}
\right)\,,
\qquad 
\omega_2=\left(
\begin{array}{r}
	1/3\\
	1/3\\
	-2/3	
\end{array}
\right) 
\,.
\eea  
The  Weyl vector  $e_0$ (the half sum of positive roots or, alternatively 
the sum of highest weights of all fundamental representations)  is equal to
\bea 
e_0=\omega_1+\omega_2\,.
\eea
The following formula for   conformal dimensions of primary fields can be readily derived from (\ref{Ln_WZW_gen})
\bea
h_{k}(\lambda)=\frac{\lambda \cdot (\lambda+2e_0)}{2(k+3)}\,,
\label{WZW_dim}
\eea
where dot is  Kronecker scalar product. By definition  primary states satisfy the conditions
\bea
\nonumber
&& H^i_0 |\phi_\lambda \rangle=\lambda_{i}|\phi_\lambda \rangle\,,
\\
\label{prst}
&& E_0^{ij}|\phi_\lambda \rangle=0\quad {\rm for} \quad i<j \,,
\\
\nonumber
&& E_n^{ij}|\phi_\lambda \rangle=H_n^{i}|\phi_\lambda \rangle=0 \quad {\rm for} \quad n>0\,.
\eea
To construct the domain wall we need also the explicit form of the $\hat{A}_2$
 modular matrix \cite{Gannon:1992ty} which in notations specified above takes the form
{\footnotesize
		\bea
	S^{(k)}_{\lambda_{n,n'},\lambda_{m,m'}}
	\equiv S^{(k)}_{\begin{psmallmatrix} n\\n'	\end{psmallmatrix}
		\begin{psmallmatrix} m\\m'	\end{psmallmatrix}}
	=\frac{i^3}{(k+3)\sqrt{3}}\sum_{w\in {\cal W}}\varepsilon(w)     
	{\rm exp}\left[
	-2\pi i\frac{w(\lambda_{m+1,m'+1})\cdot \lambda_{n+1,n'+1}}{k+3}
	\right].
	\eea
}
Here the sum runs over 6 elements of the Weyl group ${\cal W}$ acting on three component vectors through permutations.
More explicitly
{\footnotesize
\bea
\label{su3_mod_matrix}
S^{(k)}_{\begin{psmallmatrix} n\\n'	\end{psmallmatrix}\begin{psmallmatrix} m\\m'	\end{psmallmatrix}}=
\frac{-i}{\sqrt{3} (k+3)}
\bigg(
e_k\left[2 m' n'+n m'+m n'-m n\right]-e_k\left[2 m' n'+n m'+m n'+2 m n\right]
\\ \nonumber
+e_k\left[-m' n'+n m'+m n'+2 m n\right]-e_k\left[-m' n'-2 n m'+m n'-m n\right]
\\ \nonumber
-e_k\left[-m' n'+n m'-2 m n'-m n\right]+e_k\left[-m' n'-2 n m'-2 m n'-m n\right]
\bigg).
\eea
}
with notation
\bea
e_k[x]\text{:=}e^{-\frac{2 \pi  i x}{3 (k+3)}}\,.
\eea
According  to
\cite{Goddard:1984vk}
the $W_3$ minimal models can be alternatively
represented  as coset theory 
\bea
\label{W3coset}
A_2^{(k+3)}=\frac{\widehat{su}(3)_k \times \widehat{su}(3)_1}{\widehat{su}(3)_{k+1}}\,,
\eea
where   $\widehat{su}(3)_k$ stands for level $k$ WZNW model
\cite{Knizhnik:1984nr} with identification
\bea
\label{p_k_conection}
p=k+3\,.
\eea
 The stress energy tensor of the coset theory (\ref{W3coset}) is given by
\begin{equation}
T_{(su(3)_k\times su(3)_1)/su(3)_{k+1}}=T_{su(3)_k}+T_{su(3)_1}-T_{su(3)_{k+1}} \,. 
\label{diagonalcoset}
\end{equation}
For the central charge we simply get
\bea
\label{cCoset}
c_{(su(3)_k\times su(3)_1)/su(3)_{k+1}}=c_{su(3)_k}+c_{su(3)_1}-c_{su(3)_{k+1}} \,. 
\eea
Using  (\ref{KM central charge}) one easily gets convinced that 
(\ref{cCoset}) coincides with the central charge of $A_2^{(p)}$ minimal model (\ref{cp}) with already mentioned identification
(\ref{p_k_conection}).
\section{The IR/UV mixing coefficients through  domain wall construction}
\label{RGwall}
An interface 
mapping  UV observables to IR was suggested in \cite{Fredenhagen:2005an, Brunner:2007ur}.  In  \cite{Brunner:2007ur} for $N=2$ super-conformal
models  the RG domain wall was constructed  using matrix  factorization technique.
A nice algebraic construction  of RG domain walls for certain coset CFT models   was proposed in \cite{Gaiotto:2012np} and for the particular  case of Virasoro minimal models it was shown that this domain wall correctly reproduces 
perturbative results of \cite{Zamolodchikov:1987ti}. 
The proposal  was further carefully tested for various situations in
\cite{Poghosyan:2013qta,Poghossian:2013fda, Poghosyan:2014jia, Konechny:2014opa,Poghosyan:2022mfw}.

The generic  construction of \cite{Gaiotto:2012np} is also applicable for our case of interest.
Let us  briefly review this construction for our specific case.
Any  conformal
interface between two CFTs can be alternatively represented as 
a conformal boundary condition  for the direct product of these
two theories (folding trick). It is this boundary  that encodes
the UV/IR map. 
From the
coset representation of $W$ minimal models (\ref{W3coset}) 
for the direct product theory we have
\bea
\label{cos_W3}
\underbrace{\frac{\widehat{su}(3)_{k-1} \times \widehat{su}(3)_1}{\widehat{su}(3)_{k}}}_{{\rm IR : \,\, A_2^{(k+2)}}}
\times
\underbrace{\frac{\widehat{su}(3)_k \times \widehat{su}(3)_1}{\widehat{su}(3)_{k+1}}}_{\rm UV: \,\, A_2^{(k+3)}}
\sim
\frac{\widehat{su}(3)_{k-1} \times \widehat{su}(3)_1 \times \widehat{su}(3)_1}{\widehat{su}(3)_{k+1}}\,.
\eea
On the right hand side in the numerator  we get  two identical factors $\widehat{su}(3)_1$,
hence the resulting
theory admits a non trivial $\mathbb{Z}_2$ symmetry which swaps these two factors.

Using  folding trick we have \footnote{Because the RG invariant sets we consider have  fields with different holomorphic and 
	anti holomorphic parts in the next tow expressions we will label fields with two indices to distinguish them.}
\bea
\langle0|\phi^{IR}_{s\bar{r}}|RG|\phi^{UV}_{k\bar{l}}|0 \rangle =
\langle RG | \phi^{IR}_{r \bar{s}} \phi^{UV}_{k\bar{l}}  | 0 \rangle\,,
\eea
where RG denotes the interface operator.
Any one point function in  presence of a boundary
can be interpreted as  a two point function (in CFT without defect) of initial field 
and its mirror image (tilde stands for the mirror image).
\bea
\label{one_point}
\langle RG |\phi^{IR}_{r\bar{s}}   \phi^{UV}_{k\bar{l}} | 0 \rangle
\sim
\langle   \widetilde{ \phi^{IR}_s \phi^{UV}_l }| \phi^{IR}_r \phi^{UV}_k  \rangle\,.
\eea
Gaiotto's  proposal boils down to the statement that  the mirror image field can be obtained 
by simply  applying above mentioned $\mathbb{Z}_2$ operation.
This means that one should 
  find  representative a 
of corresponding state  in coset theory (\ref{cos_W3}) explicitly
and   replace   $J \leftrightarrow \tilde{J}$.
 As in standard Cardy construction these 
building blocks should be superposed with
appropriate coefficients in order to get
the ``physical" boundary state. 
Due to \cite{Gaiotto:2012np} the final
expression for the one point function in our boundary CFT is
\bea
\label{mix_coef_dom}
&\langle RG|\phi^{IR} \begin{psmallmatrix}m_i & n\\m_i' & n'	\end{psmallmatrix}\phi^{UV} \begin{psmallmatrix}n & m_j\\n' & m_j'	\end{psmallmatrix} \rangle=
{\cal M}_{ij}
\frac{\sqrt{S^{(k-1)}_{\begin{psmallmatrix} 1\\1 \end{psmallmatrix}\begin{psmallmatrix} m_i \\ m_i'	\end{psmallmatrix}} S^{(k+1)}_{\begin{psmallmatrix} 1 \\1\end{psmallmatrix} \begin{psmallmatrix} m_j \\ m_j'	\end{psmallmatrix}}}}
{S^{(k)}_{\begin{psmallmatrix} 1 \\ 1\end{psmallmatrix} \begin{psmallmatrix} n \\ n'	\end{psmallmatrix}}}\,.
\eea
where ${\cal M}_{ij}$ is properly normalized  two point function (\ref{one_point}).
After introducing some  new notations we will give precise  expressions for these quantities.
Each RG invariant set of our interest consists of  ten fields: 
\bea
\label{uv10}
&&\phi_1^{UV}=\phi^{(p)}\begin{psmallmatrix}n & n+2\\n' & n'-1	\end{psmallmatrix}\,,\quad
\phi_2^{UV}=\phi^{(p)}\begin{psmallmatrix}n & n+1\\n' & n'+1	\end{psmallmatrix}\,,\quad
\phi_3^{UV}=\phi^{(p)}\begin{psmallmatrix}n & n+1\\n' & n'-2	\end{psmallmatrix}\,,\nonumber\\
&&\phi_i^{UV}=\hat {O}_i\phi^{(p)}\begin{psmallmatrix}n & n\\n' & n'\end{psmallmatrix}\,;
\qquad {\rm for}\,\,\, i=4,5,6,7\,,\\
&&\phi_8^{UV}=\phi^{(p)}\begin{psmallmatrix}n & n-1\\n' & n'+2	\end{psmallmatrix}\,,\quad
\phi_9^{UV}=\phi^{(p)}\begin{psmallmatrix}n & n-1\\n' & n'-1	\end{psmallmatrix}\,,\quad
\phi_{10}^{UV}=\phi^{(p)}\begin{psmallmatrix}n & n-2\\n' & n'+1\end{psmallmatrix}\,,\nonumber
\eea 
where
\bea
&&\hat {O}_4:=L_{-1}\bar{L}_{-1};\quad \hat {O}_5:=L_{-1}(\bar{W}_{-1}-\frac{3w_0}{2\Delta_0}\bar{L}_{-1});\quad
\hat {O}_6:=(W_{-1}-\frac{3w_0}{2\Delta_0}L_{-1})\bar{L}_{-1};\nonumber\\ 
\label{Oi}
&&\hat {O}_7:=(W_{-1}-\frac{3w_0}{2\Delta_0}L_{-1})(\bar{W}_{-1}-\frac{3w_0}{2\Delta_0}\bar{L}_{-1})\,.
\eea 
as it was shown in \cite{Poghosyan:2022mfw}  at the IR they flow to
\bea
\label{ir10}
&&\phi_1^{IR}=\phi^{(p-1)}\begin{psmallmatrix}n-1 & n\\n'+2 & n'  \end{psmallmatrix}\,,\quad
\phi_2^{IR}=\phi^{(p-1)}\begin{psmallmatrix}n+1 & n\\n'+1 & n'	\end{psmallmatrix}\,,\quad
\phi_3^{IR}=\phi^{(p-1)}\begin{psmallmatrix}n-2 & n\\n'+1 & n'	\end{psmallmatrix}\,,\nonumber\\
&&\phi_i^{IR}=\hat {O}_i\phi^{(p-1)}\begin{psmallmatrix}n & n\\n' & n'\end{psmallmatrix}\,;
\qquad {\rm for}\,\,\, i=4,5,6,7\,,\\
&&\phi_8^{IR}=\phi^{(p-1)}\begin{psmallmatrix} n+2 & n\\n'-1 & n'	\end{psmallmatrix}\,,\quad
\phi_9^{IR}=\phi^{(p-1)}\begin{psmallmatrix}n-1 & n\\n'-1 & n'	\end{psmallmatrix}\,,\quad
\phi_{10}^{IR}=\phi^{(p-1)}\begin{psmallmatrix}n+1 & n\\n'-2 & n'	\end{psmallmatrix} \nonumber\,.
\eea
The coefficients  ${\cal M}_{ij}$ of eq.  (\ref{mix_coef_dom})  are given by
\bea
\label{Mij}
{\cal M}_{ij}=
\frac{\langle \widetilde{i_l j_l} | i_r j_r \rangle}{ \sqrt{\langle i_r j_r | i_rj_r \rangle \langle i_l j_l | i_lj_l \rangle}}\,,
\quad
i,j=1,2, \dots ,10\,,
\eea
where 
 $ | i_rj_r \rangle$ is the representative of state $  \phi^{IR}_{i_r} \phi^{UV}_{j_r} | 0 \rangle$ in direct product theory
$\widehat{su}(3)_{k-1} \times \widehat{su}(3)_1 \times \widehat{su}(3)_1$ while $|\widetilde{i_rj_r}\rangle$ is its mirror image.
We denote by $\phi_{i_r}$ ($\phi_{i_l}$)    the holomorphic (anti-holomorphic) part  of the field $\phi_i$
\footnote{The sub-indices $r$ and $l$ are essential  only when $i,j=5,6$, since the other fields have coinciding  holomorphic and anti-holomorphic parts.}.
 To identify states  $ | ij \rangle$ we have applied  the method of \cite{ Poghosyan:2014jia} 
 which is based on direct   current algebra technique.
 The details of computations and final expressions can be found in
  appendix \ref{cosetstates}. For example we got
\begin{scriptsize}
	\bea
	\label{ket11}
	|\, 1 \, 1 \, \rangle \, =\, 	\big(\, b_1 \,  (E+J)^{12}_{-1}\, +\, b_2\, \tilde{J}^{12}_{-1}\, +\, b_3\,  (E+J)^{13}_{-1}\, (E+J)^{32}_{0}\, +\, 
	b_4\,  \tilde{J}^{13}_{-1}\, (E+J)^{32}_{0}\, \big)\,  \times
	\\ \nonumber
	\big(
	a_1 E^{3 2}_{-1}+ a_2 J^{3 2}_{ -1} + a_3 E^{1 2}_{-1} E^{3 1}_ 0 + 
	a_4 J^{1 2}_{ -1} E^{3 1}_0 + a_5 H^{2}_{ -1} E^{3 2}_{ 0} + 
	a_6 J^{2 }_{ -1}  E^{3, 2}_{0} + a_7 H^{3}_{ -1}  E^{3 2}_{0} + 
	a_8 J^{3}_{ -1} E^{3 2}_{0} + 
	\\ \nonumber
	+a_9 E^{1 2}_{ -1}E^{32}_{ 0} E^{2 1}_{0} + 
	a_{10} J^{12}_{ -1} E^{32}_{0} E^{2 1}_{0} + 
	a_{11} E^{2 3}_{ -1} E^{3 2}_{ 0} E^{3 2}_{ 0} + 
	a_{12} J^{2, 3}_{ -1}E^{3, 2}_{ 0} E^{3 2}_{ 0} + 
	a_{13} E^{1 3}_{ -1}  E^{32}_{ 0} E^{3 1}_{ 0} + 
	\\ \nonumber
	+a_{14} J^{13}_{ -1} E^{3 2}_{ 0} E^{3 1}_{ 0} + 
	a_{15} E^{1 3}_{ -1} E^{3 2}_{ 0} E^{3 2}_{ 0} E^{2 1}_{ 0} + 
	a_{16} J^{1 3}_{ -1} E^{3 2}_{ 0} E^{3 2}_{ 0} E^{2 1}_{ 0}
	\big) |n-1,n'+2\rangle|1,1\rangle
	\widetilde{|1,1\rangle}\,,
	\eea
\end{scriptsize}
where  $E_n$, $J_n$ and $\tilde{J}_n$ are the  current algebra   generators of $\widehat{su}(3)_{k-1}$, $\widehat{su}(3)_1$ and $\widehat{su}(3)_1$ respectively. 
As already mentioned the expression for $| \widetilde{11} \rangle$
is obtained  by simply flipping $J_n$ and $\tilde{J_n}$ in (\ref{ket11}).
Using  Kac-Moody algebra commutation relations (\ref{KM_alg}) and properties of primary states  (\ref{prst}) in this particular case we get
\bea
\label{11}
{\cal M}_{11}=
\frac{\langle \widetilde{11} |11 \rangle}{ |\langle 11| 11 \rangle|}=\frac{1}{n+n'+1}\,.
\eea
The final result of ${\cal M}_{ij}$  can  be found in  appendix \ref{M}.
Let us denote
\bea
\langle RG |\phi^{IR}_i\phi^{UV}_j \rangle=R_{ji}\,. 
\eea
In the large $k$ limit $R$ diagonalizes the matrix of anomalous dimensions $\Gamma$. The matrix $\Gamma$ 
for our set was found in \cite{Poghosyan:2022mfw}.  We checked\footnote{To be precise the property (\ref{GmIR})
	holds after applying a  similarity transformation with diagonal matrix  $S={\rm diag}(1,1,1,1,-1,-1,1,1,1,1)$  on anomalous dimensions
	matrix found  in \cite{Poghosyan:2022mfw}.
	Such flip of signs can be  absorb  by changing the signs of  structure constants involving fields $\phi_5$ and $\phi_6$. }
that indeed
\bea
\label{GmIR}
(R^T \Gamma R)_{ij}=\Delta^{IR}_{i}\delta_{ij}\,.
\eea
Remind that  $R$ are our UV/IR mixing coefficients
\bea
\phi^{IR}_i(x)=\sum_{j=1}^{3}\phi_j(x)R_{ji}\,.
\eea
Explicit form of these mixing coefficients are given in the next section.
\section{UV/IR mixing coefficients for large $k$}
\label{Mix_coef}
The coefficients ${\cal M}_{ij}$  in eq. (\ref{mix_coef_dom})
are given in appendix \ref{M}.
Here are the  results in large  $k$ limit
\begin{scriptsize}
	\bea
&R_{11}=-\frac{\sqrt{\left(n^2+n-2\right) \left(n'^2+n'-2\right)}}{n^2 n'+n n'^2}\,,
\qquad
R_{12}=-\frac{1}{n n' (n+n')}
\sqrt{\frac{(n+2) \left(n'^2-1\right) (n+n'+2)}{(n+1) (n+n'+1)}}\,,
\eea
\bea
&R_{13}=-\frac{\sqrt{\left(n^2-4\right) \left(n'^2-1\right) \left((n+n')^2-1\right)}}{n n' (n+n')}\,,
\qquad
R_{14}=-\frac{3 (n-2) }{2 \left(n^2+n n'+n'^2-3\right)}\sqrt{\frac{(n+2) (n'-1) (n+n'+1)}{n n' (n+n')}}\,,
\eea
\bea
&R_{15}=\frac{(n-1)  (n+2 n')}{2 \left(n^2+n n'+n'^2-3\right)}
\sqrt{\frac{3 (n+2) (n'+1) (n+n'-1)}{n \left(n^2-1\right) n' (n+n')}}\,,
\eea
\bea
&R_{17}=\frac{-n^3 (n'-3)+n^2 (n' (3-5 n')+8)-n (n'-1) (n' (8 n'+5)-9)-4 n'^4+14 n'^2-18 }{2 (n+1) \left(n^2+n n'+n'^2-3\right)}
\sqrt{\frac{n+2}{n (n'-1) n' (n+n') (n+n'+1)}},
\eea
\bea
&R_{18}=\frac{-2 n n'+n-2 n'^2+2}{\left(n^2+n\right) n' (n+n')}\,,
\qquad
R_{19}=\frac{\sqrt{(n-1) (n+2) (n+n'-2) (n+n'+1)}}{n n' (n+n')}\,,
\eea
\bea
&R_{110}=\frac{1}{n n' (n+n')}\sqrt{\frac{(n+2) (n'-2) \left((n+n')^2-1\right)}{(n+1) (n'-1)}}\,,
\qquad
R_{21}=-\frac{1}{n n' (n+n')}\sqrt{\frac{\left(n^2-1\right) (n'+2) (n+n'+2)}{(n'+1) (n+n'+1)}}\,,
\eea
\bea
&R_{22}=-\frac{2 n n'+n+n'+2}{n n' (n+n') (n+n'+1)}\,,
\qquad
R_{23}=-\frac{\sqrt{(n-2) (n+1) (n+n'-1) (n+n'+2)}}{n n' (n+n')}\,,
\eea
\bea
&R_{24}=\frac{-3 (n+n'-2) }{2 \left(n^2+n n'+n'^2-3\right)}
\sqrt{\frac{(n+1) (n'+1) (n+n'+2)}{n n' (n+n')}},
\quad
R_{25}=\frac{n'-n }{2 \left(n^2+n n'+n'^2-3\right)}
\sqrt{\frac{3 (n-1) (n'-1) (n+n'-1) (n+n'+2)}{n n' (n+n') (n+n'+1)}},
\eea
\bea
&R_{27}=\frac{\left(-\left((n+3) n'^3\right)+2 (n-4) (n+1) n'^2-n (n (n+6)+2) n'+n (9-n (3 n+8))+9 (n'+2)\right)}{2 (n+n'+1) \left(n^2+n n'+n'^2-3\right)}
 \sqrt{\frac{n+n'+2}{n (n+1) n' (n'+1) (n+n')}},
 \eea
 \bea
 &R_{29}=-\frac{\sqrt{\left(n^2-1\right) \left(n'^2-1\right) \left((n+n')^2-4\right)}}{n n' (n+n')}\,,
 \qquad
 R_{210}=-\frac{\sqrt{(n'-2) (n'+1) (n+n'-1) (n+n'+2)}}{n n' (n+n')}\,,
 \eea
 \bea
 &R_{31}=-\frac{\sqrt{\left(n^2-1\right) \left(n'^2-4\right) \left((n+n')^2-1\right)}}{n n' (n+n')}\,,
 \qquad
 R_{33}=\frac{\sqrt{(n-2) (n+1) (n'-2) (n'+1)}}{n n' (n+n')}\,,
 \eea
 \bea
 &R_{34}=\frac{3 (n'+2) }{2 \left(n^2+n n'+n'^2-3\right)}
 \sqrt{\frac{(n+1) (n'-2) (n+n'-1)}{n n' (n+n')}},
 \quad
 R_{35}=\frac{(n-1) (n'+1) (2 n+n') }{2 \left(n^2+n n'+n'^2-3\right)}
 \sqrt{\frac{3 (n'-2) (n+n'+1)}{n (n-1) n' \left(n'^2-1\right) (n+n')}},
 \eea
 \bea
 &R_{37}=\frac{8 n^3 n'+n^2 (n'+2) (5 n'-7)+4 n^4+n n' (n' (n'+3)-14)+(n'-3) \left(3 n'^2+n'-6\right) }{2 (n'-1) \left(n^2+n n'+n'^2-3\right)}
 \sqrt{\frac{n'-2}{n (n+1) n' (n+n'-1) (n+n')}},
 \eea
 \bea
 &R_{39}=-\frac{1}{n n' (n+n')}
 \sqrt{\frac{\left(n^2-1\right) (n'-2) (n+n'-2)}{(n'-1) (n+n'-1)}}\,,
 \qquad
 R_{310}=-\frac{2 n^2+2 n n'+n'-2}{n (n'-1) n' (n+n')}\,,
 \eea
 \bea
 &R_{41}=-\frac{3 (n'-2) }{2 \left(n^2+n n'+n'^2-3\right)}
 \sqrt{\frac{(n-1) (n'+2) (n+n'+1)}{n n' (n+n')}},
 \quad
 R_{43}=\frac{3 (n+2) }{2 \left(n^2+n n'+n'^2-3\right)}
 \sqrt{\frac{(n-2) (n'+1) (n+n'-1)}{n n' (n+n')}},
 \eea
 \bea
 &R_{45}=0\,,
 \quad
 R_{47}=\frac{9}{2 \left(n^2+nn'+n'^2-3\right)}\,,
 \quad
 R_{49}=\frac{3 (n+n'+2) }{2 \left(n^2+nn'+n'^2-3\right)}
 \sqrt{\frac{(n-1) (n'-1) (n+n'-2)}{nn' (n+n')}}\,,
 \eea
 \bea
 &R_{410}=\frac{3 (n'+2) }{2 \left(n^2+n n'+n'^2-3\right)}
 \sqrt{\frac{(n+1) (n'-2) (n+n'-1)}{n n' (n+n')}},
 \quad
 R_{51}=\frac{ (n+1) (1-n') (2 n+n') }{2 \left(n^2+n n'+n'^2-3\right)}
 \sqrt{\frac{3(n'+2) (n+n'-1)}{n \left(n'^3-n'\right) \left(n^2+n n'+n+n'\right)}},
 \eea
 \bea
 &R_{53}=\frac{-(n+1) (n+2 n') }{2 \left(n^2+n n'+n'^2-3\right)}
 \sqrt{\frac{3 (n-2) (n'-1) (n+n'+1)}{n n' \left(n^2-1\right)  (n+n')}},
 \,\,
 R_{57}=\frac{\sqrt{3} (n'-n) (2 n+n') (n+2 n')}{\sqrt{\left(n^2-1\right) \left(n'^2-1\right) (n+n'-1) (n+n'+1)} \left(n^2+n n'+n'^2-3\right)},
 \eea
 \bea
 &R_{59}=\frac{ (n+1) (n'+1) (n-n') }{2 \left(n^2+n n'+n'^2-3\right)}
 \sqrt{\frac{3(n+n'-2) (n+n'+1)}{n (n+1) n' (n'+1) (n+n'-1) (n+n')}}\,,
 \eea
 \bea
 &R_{71}=\frac{-8 n^3 n'-n^2 (n'-2) (5 n'+7)-4 n^4+n n' (14-(n'-3) n')+(n'+3) (n' (3 n'-1)-6) }{2 (n'+1) \left(n^2+n n'+n'^2-3\right)}
 \sqrt{\frac{n'+2}{n (n-1) n' (n+n') (n+n'+1)}},
 \eea
 \bea
 &R_{73}=\frac{\left(n^3 (n'+3)+n^2 (n'-1) (5 n'+8)+n (n'+1) (n' (8 n'-5)-9)+2 \left(2 n'^4-7 n'^2+9\right)\right) }{2 (n-1) \left(n^2+n n'+n'^2-3\right)}
 \sqrt{\frac{n-2}{n n' (n'+1) (n+n'-1) (n+n')}},
 \eea
 \bea
 R_{77}&=&-\frac{27}{2 \left(n^2+n n'+n'^2-3\right)}+\frac{5-n^2}{n^2-1}+\frac{2 (n-1) n-1}{(n-1) n (n+n'-1)}
 \\ \nonumber
 &+&\frac{2 n (n+1)-1}{n (n+1) (n'-1)}+\frac{1-2 (n-1) n}{(n-1) n (n'+1)}+\frac{1-2 n (n+1)}{n (n+1) (n+n'+1)},
 \eea
 \bea
 &R_{79}=\frac{n^3 (n'-3)-2 n^2 (n'-1) (n'+4)+n (n'+1) ((n'-7) n'+9)-(n'-3) \left(3 n'^2+n'-6\right) }{2 (n+n'-1) \left(n^2+n n'+n'^2-3\right)}\sqrt{\frac{n+n'-2}{(n-1) n (n'-1) n' (n+n')}},
 \\
 &R_{81}=\frac{-2 n (n+n')+n'+2}{n n' (n'+1) (n+n')}\,,
 \qquad
 R_{83}=\frac{1}{n n' (n+n')}
 \sqrt{\frac{(n-2) (n'+2) \left((n+n')^2-1\right)}{(n-1) (n'+1)}}\,,
 \eea
 \bea
 &R_{89}=\frac{\sqrt{(n'-1) (n'+2) (n+n'-2) (n+n'+1)}}{n n' (n+n')}\,,
 \qquad
 R_{93}=-\frac{1}{n n' (n+n')}\sqrt{\frac{(n-2) \left(n'^2-1\right) (n+n'-2)}{(n-1) (n+n'-1)}}\,,
 \eea
 \bea
 &R_{99}=\frac{-2 n n'+n+n'-2}{n n' (n+n'-1) (n+n')}\,,
 \qquad
 R_{103}=-\frac{2 n n'+n+2 n'^2-2}{(n-1) n n' (n+n')}\,,
 \eea
\end{scriptsize}
The  remaining coefficients are related to those mentioned above as 
\begin{scriptsize}
	\bea
		R_{15}=R_{16}=R_{58}=R_{68},
		\,\,
		R_{25}=R_{26}=R_{52}=R_{62},
			\,\,
		R_{35}=R_{36}=R_{510}=R_{610},
		\,\,
			R_{44}=R_{55}=R_{66},
			\\ \nonumber
			R_{45}=R_{46}=R_{54}=R_{64}\,,
			\quad
		R_{47}=R_{56}=R_{65}=R_{74}\,,
		\quad
	    R_{51}=R_{61}=R_{85}=R_{86}\,,
		\\ \nonumber
	    R_{53}=R_{63}=R_{105}=R_{106}\,,
		\quad
	    R_{57}=R_{67}=R_{75}=R_{76}\,,
		\quad
	     R_{59}=R_{69}=R_{95}=R_{96}\,,
\eea
\bea
	R_{11}=R_{88},
	\quad
	R_{12}=R_{28},
	\quad
	R_{13}=R_{10 8},
	\quad
	R_{14}=R_{48},
	\quad
	R_{17}=R_{78},
	\quad
	R_{19}=R_{98},
	\quad
	R_{110}=R_{38},
	\\ \nonumber
	R_{21}=R_{82},
	\quad
	R_{23}=R_{102}
	\quad
	R_{24}=R_{42}
	\quad
	R_{27}=R_{72}
	\quad
	R_{29}=R_{92}
	\quad
	R_{210}=R_{32}
	\quad
	R_{31}=R_{810}
	\\ \nonumber
	R_{33}=R_{1010},
	\quad
	R_{34}=R_{410},
	\quad
    R_{37}=R_{710},
	\quad
	R_{39}=R_{910},
	\quad
	R_{41}=R_{84},
	\quad
	R_{43}=R_{104},
	\\ \nonumber
    R_{49}=R_{94},
    \,\,
    R_{71}=R_{87},
    \,\,
    R_{73}=R_{107},
    \,\,
    R_{79}=R_{97},
    \,\,
    R_{83}=R_{101},
    \,\,
    R_{89}=R_{91},
    \,\,
    R_{93}=R_{109}.
    \eea
\end{scriptsize}
\section{The  matrix of anomalous W-weights}
Since we have already computed the mixing matrix $R$ we can derive   
the matrix of anomalous W-weights in IR limit using
\bea
\label{W}
\mathfrak{Q}_{\alpha\beta}^{IR}=\left(R\, w^{IR} R^T\right)_{\alpha\beta}\,.
\eea
It was argued in  \cite{Poghosyan:2022mfw} that 
the leading order formula for anomalous W-weights in perturbed CFT for primary fields
takes the form
\bea
\label{QmatEl}
\mathfrak{Q}_{\alpha\beta}=w_{\beta}\delta_{\alpha\beta}
+\pi g a_{\alpha\beta} C_{\,\,\beta}^\alpha+O(g^2)\,.
\eea
The case of our interest is more complicated since
the RG invariant set involves four secondary fields.
Using (\ref{W}) and (\ref{W}) we have cheeked that the coefficients $a_{\alpha\beta}$, with $\alpha,\beta=1,2,3,8,9,10$ corresponding 
to primaries are given by 
\bea
a_{\alpha \beta}=\frac{w_{\alpha}-w_{\beta}}{\Delta_{\alpha}-\Delta_{\beta}}
\quad 
{\rm for}
\quad
\alpha \ne \beta
\eea
as suggested in \cite{Poghosyan:2022mfw}. For the diagonal elements we get the expressions
\bea
a_{11}=\frac{\sqrt{\frac{3}{2}} (n+2) (n+2 n')}{2 (n+3) n'^2+2 n (n+3) n'-n (n+2)}\,,
\eea
\bea
a_{22}=-\frac{\sqrt{\frac{3}{2}} (n-n') (n+n'+2)}{n^2 (2 n'+1)+2 n (n' (n'+4)+1)+n' (n'+2)}\,,
\eea
\bea
a_{33}=-\frac{\sqrt{\frac{3}{2}} (n'-2) (2 n+n')}{2 n^2 (n'-3)+2 n (n'-3) n'+(n'-2) n'}\,,
\eea
\bea
a_{88}=-\frac{\sqrt{\frac{3}{2}} (n'+2) (2 n+n')}{2 n^2 (n'+3)+2 n n' (n'+3)-n' (n'+2)}\,,
\eea
\bea
a_{99}=-\frac{\sqrt{\frac{3}{2}} (n-n') (n+n'-2)}{n^2 (2 n'-1)+2 n ((n'-4) n'+1)-(n'-2) n'}\,,
\eea
\bea
a_{1010}=\frac{\sqrt{\frac{3}{2}} (n-2) (n+2 n')}{2 (n-3) n'^2+2 (n-3) n n'+(n-2) n}\,.
\eea
We were able to recover the same values by using perturbation theory
closely following the route described in \cite{Poghosyan:2022mfw}. Since the remaining secondary fields 
do not have definite $W_0$ eigenvalues, it is natural to expect that the expression (\ref{QmatEl}) should 
be generalized as
\bea
\label{QmatEldes}
\mathfrak{Q}_{\alpha\beta}=w_{\alpha\beta}
+\pi g a_{\alpha\beta} C_{\,\,\beta}^\alpha+O(g^2)\,.
\eea  
where 
\bea
W_{0}\phi_{\alpha}=w_{\alpha \beta}\phi_{\beta}\,.
\eea   
Explicitly 
\bea
& W_0\phi_4 =\frac{(\Delta_0+3)w_0}{\Delta_0} \Phi_4+\sqrt{2\left(\frac{ (2-c+32  \Delta_0)}{22+5c}-\frac{9 w_0^2}{2\Delta_0^2}\right)} \phi_6  \,,
\\
& W_0 \phi_5 = \frac{(\Delta_0+3)w_0}{\Delta_0} \Phi_5+ \sqrt{2\left(\frac{ (2-c+32  \Delta_0)}{22+5c}-\frac{9 w_0^2}{2\Delta_0^2}\right)} \phi_7  \,,
\\
& W_0 \phi_6 =\frac{(\Delta_0-3)  w_0}{\Delta_0}\Phi_6+ \sqrt{2\left(\frac{ (2-c+32  \Delta_0)}{22+5c}-\frac{9 w_0^2}{2\Delta_0^2}\right)}\phi_4 \,,
\\
& W_0 \phi_7=\frac{(\Delta_0-3)  w_0}{\Delta_0}\Phi_7+  \sqrt{2\left(\frac{ (2-c+32  \Delta_0)}{22+5c}-\frac{9 w_0^2}{2\Delta_0^2}\right)}\phi_5\,.
\eea
From (\ref{W}) and(\ref{QmatEldes})
 we know that
 \bea
 a_{\alpha \beta}=\frac{\mathfrak{Q}_{\alpha\beta}^{IR}-w^{UV}_{\alpha \beta}}{\pi g C_{\,\,\beta}^\alpha}\,.
 \eea
 Thus we have prediction for all $a_{\alpha \beta}$ coming from domain wall approach.
 It would be interesting, though not easy, to check this results using perturbation theory also for
 the indices $\alpha,\beta$ corresponding to descendants.  

\section*{Acknowledgments}
The  work of H.P. was supported by Armenian SCS grants 21AG‐1C060, 20TTWS-1C035,
and ANCEF 22AN:PS-mathph-2697. A.P. acknowledge the support in the framework of Armenian SCS
grants 20TTWS-1C035, 21AG‐1C062.
\begin{appendix}
\section{Field identification  in  coset theory}
\label{cosetstates}
Let     $|ij\rangle$ be the representation of
$ \phi^{IR}_{i} \phi^{UV}_j  | 0 \rangle$  in 
$\widehat{su}(3)_{k-1} \times \widehat{su}(3)_1 \times \widehat{su}(3)_1$.
Remind that 
\bea
\label{cos_W3_ap}
\underbrace{\frac{\widehat{su}(3)_{k-1} \times \widehat{su}(3)_1}{\widehat{su}(3)_{k}}}_{{\rm IR : \,\, A_2^{(k+2)}}}
\times
\underbrace{\frac{\widehat{su}(3)_k \times \widehat{su}(3)_1}{\widehat{su}(3)_{k+1}}}_{\rm UV: \,\, A_2^{(k+3)}}
\sim
\frac{\overbrace{\widehat{su}(3)_{k-1}}^{E} \times \overbrace{\widehat{su}(3)_1}^{J} \times \overbrace{\widehat{su}(3)_1}^{\tilde{J}}}
{\underbrace{\widehat{su}(3)_{k+1}}_{E+J+\tilde{J}}}\,.
\eea
To find  $|ij\rangle$   we first represent   $|\phi_i^{IR}\rangle $ in $\widehat{su}(3)_{k-1} \times \widehat{su}(3)_1$ and then
 $|\phi_j^{UV}\rangle $ in $\widehat{su}(3)_k \times \widehat{su}(3)_1$. 
Below we will give details of computation only for the case  $|11\rangle$, since treatment of other cases is quite similar.

Lets analise  the IR primary $\phi_1^{IR}=\phi^{(k+2)}\begin{psmallmatrix}n-1 & n\\n'+2 & n'  \end{psmallmatrix}$.
From (\ref{dim}) and (\ref{WZW_dim}) one can check that
\bea
\label{h11}
	\Delta ^{(k+2)}
\begin{psmallmatrix}
	n-1 & n\\
	n'+2 & n'	
\end{psmallmatrix}= 	
h_{k-1}(n-1,n'+2)+h_{1}(1,1)+1-h_{k}(n,n')\,.
\eea
Notice also (see (\ref{lambda}))
\bea
\lambda_{n,n'}=
\lambda_{n-1,n'+2}+\lambda_{1,1}+\left(
\begin{array}{r}
	0\\
	-1\\
	1
\end{array}
\right)\,.
\eea
We are looking for a level 1 (due to the shift by $+1$ in (\ref{h11})) state inside module generated by 
$ |n-1,n'+2\rangle|1,1\rangle$ in $\widehat{su}(3)_{k-1} \times \widehat{su}(3)_1$
with total weight 
\bea
\nonumber
\lambda_{n-1,n'+2}+\lambda_{1,1}+\left(
\begin{array}{r}
	0\\
	-1\\
	1
\end{array}
\right)
\eea
which, in addition, is a primary with respect to $E+\tilde{J}$.
A  generic level $1$ state with required weight  can be written as
\begin{scriptsize}
	\bea
	\label{IR1}
	|\phi^{IR}_1\rangle=
	\bigg(
	a_1 E^{3 2}_{-1}+ a_2 J^{3 2}_{ -1} + a_3 E^{1 2}_{-1} E^{3 1}_ 0 + 
	a_4 J^{1 2}_{ -1} E^{3 1}_0 + a_5 H^{2}_{ -1} E^{3 2}_{ 0} + 
	a_6 J^{2 }_{ -1}  E^{3, 2}_{ 0} + a_7 H^{3}_{ -1}  E^{3 2}_{ 0} + 
	a_8 J^{3}_{ -1} E^{3 2}_{ 0} + 
	\\ \nonumber
	+a_9 E^{1 2}_{ -1}E^{32}_{ 0} E^{2 1}_{0} + 
	a_{10} J^{12}_{ -1} E^{32}_{0} E^{2 1}_{0} + 
	a_{11} E^{2 3}_{ -1} E^{3 2}_{ 0} E^{3 2}_{ 0} + 
	a_{12} J^{2, 3}_{ -1}E^{3, 2}_{ 0} E^{3 2}_{ 0} + 
	a_{13} E^{1 3}_{ -1}  E^{32}_{ 0} E^{3 1}_{ 0} + 
	\\ \nonumber
	+a_{14} J^{13}_{ -1} E^{3 2}_{ 0} E^{3 1}_{ 0} + 
	a_{15} E^{1 3}_{ -1} E^{3 2}_{ 0} E^{3 2}_{ 0} E^{2 1}_{ 0} + 
	a_{16} J^{1 3}_{ -1} E^{3 2}_{ 0} E^{3 2}_{ 0} E^{2 1}_{ 0}
	\bigg) |n-1,n'+2\rangle|1,1\rangle\,.
	\eea
\end{scriptsize}
To single out the correct  representative it remains to impose
\bea
\left(E^{ij}_n+J^{ij}_n\right)|\phi^{IR}_1\rangle=0\,,
\quad
{\rm for}
\quad
n>0\,,
\\
\left(E^{ij}_0+J^{ij}_0\right)|\phi^{IR}_1\rangle=0\,,
\quad
{\rm for}
\quad
i<j\,,
\eea
which specifies all coefficients $a_i$  
up to a common factor. Setting $a_1=1$ we have found that
\begin{scriptsize}
\bea
a_1\equiv1\,, 
\quad
a_2= - (k+n'+3)\,,
\quad
a_3= -\frac{ (n'+2)}{(n'+1) (n+n')}\,,
\quad
a_4= \frac{ (n'+2) (k+n'+3)}{(n'+1) (n+n')}\,,
\\ \nonumber
a_5= -\frac{1}{n'+1}\,,
\quad
a_6= \frac{ (k+n'+3)}{n'+1}\,,
\quad
a_7= \frac{1}{n'+1}\,,
\quad
a_8= -\frac{ (k+n'+3)}{n'+1}\,,
\quad
a_9= \frac{1}{(n'+1) (n+n')}\,,
\\ \nonumber
a_{10}= -\frac{ (k+n'+3)}{(n'+1) (n+n')}\,,
\quad
a_{11}= -\frac{1}{n'^2+n'}\,,
\quad
a_{12}= \frac{ (k+n'+3)}{n' (n'+1)}\,,
\quad
a_{13}= -\frac{ (n'+2)}{n' (n'+1) (n+n')}\,,
\\ \nonumber
a_{14}= \frac{ (n'+2) (k+n'+3)}{n' (n'+1) (n+n')}\,,
\quad
a_{15}= \frac{1}{\left(n'^2+n'\right) (n+n')}\,,
\quad
a_{16}= -\frac{ (k+n'+3)}{n' (n'+1) (n+n')}\,.
\eea
\end{scriptsize}
A similar  analyses  is done for the  $\phi_1^{UV}=\phi^{(p)}\begin{psmallmatrix}n & n+2\\n' & n'-1	\end{psmallmatrix}$.
Here we have
\bea
\Delta ^{(k+3)}
\begin{psmallmatrix}
	n & n+2\\
	n' & n'	-1
\end{psmallmatrix}= 	
h_{k}(n,n')+h_{1}(1,1)-h_{k+1}(n+2,n'-1)+1
\eea
and 
\bea
\lambda_{n+2,n'-1}=\lambda_{n,n'}+\lambda_{1,1}+\left(
\begin{array}{r}	1\\	-1\\    0
\end{array}
\right)\,.
\eea
So
\bea
\label{UV1}
|\phi^{UV}_1\rangle=
\left(b_1 M^{12}_{-1}+b_2 \tilde{J}^{12}_{-1}+b_3 M^{13}_{-1}M^{32}_{0}+b_4 \tilde{J}^{13}_{-1}M^{32}_{0}\right)|n,n'\rangle|1,1\rangle\,.
\eea
Now impose the primality condition
\bea
\left(M^{ij}_n+\tilde{J}^{ij}_n\right)|\phi^{UV}_1\rangle=0\,,
\quad
{\rm for}
\quad
n>0\,,
\\
\left(M^{ij}_0+\tilde{J}^{ij}_0\right)|\phi^{UV}_1\rangle=0\,,
\quad
{\rm for}
\quad
i<j\,.
\eea
where $M=E+J$.
This leads to
\begin{small}
	\bea
	b_1=1,
	\quad
	b_2 = n -2 - k ,
	\quad 
	b_3 =\frac{1}{n'-1 }, 
	\quad
	b_4 =\frac  {n-2 - k }{n'-1 }\,.
	\eea
\end{small}
From (\ref{cos_W3_ap}), (\ref{IR1}) and (\ref{UV1}) for   $|11\rangle$ we finally obtain
\begin{small}
	\bea
	\label{11st}
	|11\rangle=
	\big(b_1 (E+J)^{12}_{-1}+b_2\tilde{J}^{12}_{-1}&+&b_3 (E+J)^{13}_{-1}(E+J)^{32}_{0}+
	\\ \nonumber
	&+&b_4 \tilde{J}^{13}_{-1}(E+J)^{32}_{0}\big)
	|\phi^{IR}_1\rangle\widetilde{|1,1\rangle}\,.
	\eea
\end{small}
Other primary fields can be presented as  coset  using similar methods. Here are the results
for IR fields
\begin{scriptsize}
\bea
\label{phiIR1}
|\phi^{IR}_1\rangle = \big(
 E_{-1}^{3 2}+a^1_{2} J_{-1}^{3 2}+a^1_{3} E_{-1}^{1 2}E_0^{3 1}+a^1_{4} J_{-1}^{1 2}E_0^{3 1}+a^1_{5} E_{-1}^{2 2}E_0^{3 2}+a^1_{6} J_{-1}^{2 2}E_0^{3 2}+a^1_{7} E_{-1}^{3 3}E_0^{3 2}+a^1_{8} J_{-1}^{3 3}E_0^{3 2}
\nonumber \\
+a^1_{9} E_{-1}^{1 2}E_0^{3 2}E_0^{2 1}+a^1_{10} J_{-1}^{1 2}E_0^{3 2}E_0^{2 1}+a^1_{11} E_{-1}^{2 3}E_0^{3 2}E_0^{3 2}+a^1_{12} J_{-1}^{2 3}E_0^{3 2}E_0^{3 2}+a^1_{13} E_{-1}^{1 3}E_0^{3 2}E_0^{3 1}
\\ \nonumber
+a^1_{14} J_{-1}^{1 3}E_0^{3 2}E_0^{3 1}+a^1_{15} E_{-1}^{1 3}E_0^{3 2}E_0^{3 2}E_0^{2 1}+a^1_{16} J_{-1}^{1 3}E_0^{3 2}E_0^{3 2}E_0^{2 1}\big)	|n-1,n'+2\rangle|1,1\rangle\;,
\eea
\bea
\nonumber
|\phi^{IR}_2\rangle = \big(
 E_{-1}^{3 1}+a^2_{2} J_{-1}^{3 1}+a^2_{3} E_{-1}^{3 3}E_0^{3 1}+a^2_{4} J_{-1}^{3 3}E_0^{3 1}+a^2_{5} E_{-1}^{1 1}E_0^{3 1}+a^2_{6} J_{-1}^{1 1}E_0^{3 1}+a^2_{7} E_{-1}^{3 2}E_0^{2 1}
  +a^2_{8} J_{-1}^{3 2}E_0^{2 1} 
  \\ \nonumber
+a^2_{9} E_{-1}^{2 1}E_0^{3 2} +a^2_{10} J_{-1}^{2 1}E_0^{3 2}+a^2_{11} E_{-1}^{1 1}E_0^{3 2}E_0^{2 1}+a^2_{12} J_{-1}^{1 1}E_0^{3 2}E_0^{2 1}+a^2_{13} E_{-1}^{3 3}E_0^{3 2}E_0^{2 1}+
 a^2_{14} J_{-1}^{3 3}E_0^{3 2}E_0^{2 1}
 \\\nonumber
 +a^2_{15} E_{-1}^{1 2}E_0^{3 1}E_0^{2 1}+a^2_{16} J_{-1}^{1 2}E_0^{3 1}E_0^{2 1}+a^2_{17} E_{-1}^{2 3}E_0^{3 2}E_0^{3 1}+a^2_{18} J_{-1}^{2 3}E_0^{3 2}E_0^{3 1}+a^2_{19} E_{-1}^{1 3}E_0^{3 1}E_0^{3 1}
 +a^2_{20} J_{-1}^{1 3}E_0^{3 1}E_0^{3 1}
 \\\nonumber
+a^2_{21} E_{-1}^{1 2}E_0^{3 2}E_0^{2 1}E_0^{2 1} +a^2_{22} J_{-1}^{1 2}E_0^{3 2}E_0^{2 1}E_0^{2 1}+a^2_{23} E_{-1}^{1 3}E_0^{3 2}E_0^{3 1}E_0^{2 1}+a^2_{24} J_{-1}^{1 3}E_0^{3 2}E_0^{3 1}E_0^{2 1}+a^2_{25} E_{-1}^{2 3}E_0^{3 2}E_0^{3 2}E_0^{2 1}
 \\
 +a^2_{26} J_{-1}^{2 3}E_0^{3 2}E_0^{3 2}E_0^{2 1}+a^2_{27} E_{-1}^{1 3}E_0^{3 2}E_0^{3 2}E_0^{2 1}E_0^{2 1}+a^2_{28} J_{-1}^{1 3}E_0^{3 2}E_0^{3 2}E_0^{2 1}E_0^{2 1}\big)
 	|n+1,n'+1\rangle|1,1\rangle\,,
 	\qquad
\eea
\bea
|\phi^{IR}_3\rangle = 
\big( E_{-1}^{1 2}+a^3_{2} J_{-1}^{1 2}+a^3_{3} E_{-1}^{1 3}E_0^{3 2}+a^3_{4} J_{-1}^{1 3}E_0^{3 2}\big)	|n-2,n'+1\rangle |1,1\rangle\,,
\eea
\bea
|\phi^{IR}_8\rangle = \big(
 E_{-1}^{2 1}+a^8_{2} J_{-1}^{2 1}+a^8_{3} E_{-1}^{1 1}E_0^{2 1}+a^8_{4} J_{-1}^{1 1}E_0^{2 1}+a^8_{5} E_{-1}^{2 2}E_0^{2 1}+a^8_{6} J_{-1}^{2 2}E_0^{2 1}
+a^8_{7} E_{-1}^{2 3}E_0^{3 1}+a^8_{8} J_{-1}^{2 3}E_0^{3 1}
\nonumber \\ \nonumber
+a^8_{9} E_{-1}^{1 2}E_0^{2 1}E_0^{2 1}
+a^8_{10} J_{-1}^{1 2}E_0^{2 1}E_0^{2 1}+a^8_{11} E_{-1}^{1 3}E_0^{3 1}E_0^{2 1}+a^8_{12} J_{-1}^{1 3}E_0^{3 1}E_0^{2 1}+a^8_{13} E_{-1}^{2 3}E_0^{3 2}E_0^{2 1}
\\
+a^8_{14} J_{-1}^{2 3}E_0^{3 2}E_0^{2 1}
+a^8_{15} E_{-1}^{1 3}E_0^{3 2}E_0^{2 1}E_0^{2 1}+a^8_{16} J_{-1}^{1 3}E_0^{3 2}E_0^{2 1}E_0^{2 1}\big)|n+2,n'-1\rangle |1,1\rangle\,,
\eea
\bea
|\phi^{IR}_9\rangle = \big( a^9_{1} E_{-1}^{1 3}+a^9_{2} J_{-1}^{1 3} \big)|n-1,n'-1\rangle |1,1\rangle\,,
\eea
\bea
\label{phiIR10}
|\phi^{IR}_{10}\rangle = \big(
 E_{-1}^{2 3}+a^{10}_{2} J_{-1}^{2 3}+a^{10}_{3} E_{-1}^{1 3}E_0^{2 1}+a^{10}_{4} J_{-1}^{1 3}E_0^{2 1}
  \big) |n+1,n'-2\rangle |1,1\rangle\,,
\eea
\end{scriptsize}
where the coefficients are
\begin{scriptsize}
	\bea
	&a^{(1)}_{2}= - (k+n'+3), \; a^{(1)}_{3}= \frac{- (n'+2)}{(n'+1) (n+n')}, 
	\; a^{(1)}_{4}= \frac{ (n'+2) (k+n'+3)}{(n'+1) (n+n')}, \; a^{(1)}_{5}= \frac{-1}{n'+1},\;
	a^{(1)}_{6}= \frac{k+n'+3}{n'+1},
	\nonumber
	\\ \nonumber
	&  a^{(1)}_{7}= \frac{1}{n'+1}, \; a^{(1)}_{8}= -\frac{k+n'+3}{n'+1}, \; a^{(1)}_{9}= \frac{1}{(n'+1) (n+n')}, 
	\; a^{(1)}_{10}= \frac{- (k+n'+3)}{(n'+1) (n+n')}, \; a^{(1)}_{11}= \frac{-1}{n'^2+n'},
	\; a^{(1)}_{12}= \frac{k+n'+3}{n' (n'+1)},
	\\ 
	& 
	a^{(1)}_{13}= -\frac{n'+2}{n' (n'+1) (n+n')}, \; a^{(1)}_{14}= \frac{ (n'+2) (k+n'+3)}{n' (n'+1) (n+n')}, 
	\;
	a^{(1)}_{15}= \frac{1}{\left(n'^2+n'\right) (n+n')}, \; a^{(1)}_{16}= -\frac{k+n'+3}{n' (n'+1) (n+n')}\,,
	\eea 
	\bea
	&a^{(2)}_{2}= -(k+n+n'+3), \;a^{(2)}_{3}= \frac{n'-1}{n' (n+n'+1)}, \;a^{(2)}_{4}= -\frac{(n'-1) (k+n+n'+3)}{n' (n+n'+1)}, \; a^{(2)}_{5}= -\frac{n'+2}{n n'+n'^2+n'}, 
	\\ \nonumber
	&a^{(2)}_{6}= \frac{(n'+2) (k+n+n'+3)}{n' (n+n'+1)}, \; a^{(2)}_{7}= \frac{1}{n}, \; a^{(2)}_{8}= -\frac{k+n+n'+3}{n}, \; a^{(2)}_{9}= -\frac{1}{n'}, \; a^{(2)}_{10}= \frac{k+n+n'+3}{n'}, \; a^{(2)}_{11}= \frac{2 n+n'}{n n' (n+n'+1)},
	\\ \nonumber
	& \;a^{(2)}_{12}= -\frac{(2 n+n') (k+n+n'+3)}{n n' (n+n'+1)}, \; a^{(2)}_{13}= \frac{n+2 n'}{n n' (n+n'+1)}, \; a^{(2)}_{14}= -\frac{(n+2 n') (k+n+n'+3)}{n n' (n+n'+1)}, \;a^{(2)}_{15}= -\frac{n'+2}{n n' (n+n'+1)}, 
	\\ \nonumber
	&a^{(2)}_{16}= \frac{(n'+2) (k+n+n'+3)}{n n' (n+n'+1)}, \; a^{(2)}_{17}= \frac{-1}{n n'+n'^2+n'}, \; a^{(2)}_{18}= \frac{k+n+n'+3}{n n'+n'^2+n'}, \; a^{(2)}_{19}= \frac{-(n'+1)}{n' (n+n') (n+n'+1)}, \;a^{(2)}_{20}= \frac{(n'+1) (k+n+n'+3)}{n' (n+n') (n+n'+1)}, 
	\\ \nonumber
	&a^{(2)}_{21}= \frac{1}{n n' (n+n'+1)}, \; a^{(2)}_{22}= -\frac{k+n+n'+3}{n n' (n+n'+1)}, \; a^{(2)}_{23}= \frac{n-n'-2}{n n' (n+n') (n+n'+1)}, \; a^{(2)}_{24}= \frac{-(n-n'-2) (k+n+n'+3)}{n n' (n+n') (n+n'+1)}, 
	\\ \nonumber
	&a^{(2)}_{25}= \frac{-1}{n n' (n+n'+1)}, \; a^{(2)}_{26}= \frac{k+n+n'+3}{n n' (n+n'+1)}, \; a^{(2)}_{27}= \frac{1}{n n' (n+n') (n+n'+1)}, \; a^{(2)}_{28}= \frac{-(k+n+n'+3)}{n n' (n+n') (n+n'+1)},
	\eea
	\bea
	& a^{(3)}_{2}= n-k-3,\quad a^{(3)}_3= \frac{1}{n'}, \quad a^{(3)}_4= \frac{n-k-3}{n'},
	\eea
	\bea
	\nonumber
	& a^{(8)}_{2}= - (k+n+3), \;a^{(8)}_{3}= -\frac{1}{n+1}, \; a^{(8)}_{4}= \frac{k+n+3}{n+1}, \; a^{(8)}_{5}= \frac{1}{n+1}, \; a^{(8)}_{6}= -\frac{k+n+3}{n+1}, \; a^{(8)}_{7}= \frac{1}{n+n'}, 
	\\
	\nonumber
	& a^{(8)}_{8}= -\frac{k+n+3}{n+n'}, \; a^{(8)}_{9}= -\frac{1}{n^2+n}, \; a^{(8)}_{10}= \frac{k+n+3}{n (n+1)}, \; a^{(8)}_{11}= -\frac{1}{(n+1) (n+n')}, \; a^{(8)}_{12}= \frac{ k+n+3}{(n+1) (n+n')}, 
	\\
	& a^{(8)}_{13}= \frac{1}{(n+1) (n+n')}, \; a^{(8)}_{14}= -\frac{k+n+3}{(n+1) (n+n')}, \; a^{(8)}_{15}= -\frac{1}{\left(n^2+n\right) (n+n')}, \; a^{(8)}_{16}= \frac{ k+n+3}{n (n+1) (n+n')},
	\eea
	\bea
	a^{(9)}_2= - (k-n-n'+3),
	\eea
	\bea
	&a^{(10)}_{2}=  n'-k-3,\quad a^{(10)}_{3}= -\frac{1}{n},\quad a^{(10)}_{4}= \frac{k-n'+3}{n}\,.
	\eea
\end{scriptsize}
And the resulting expression  for UV states read
\begin{scriptsize}
	\bea
	\label{phiUV1}
	|\phi^{UV}_1\rangle = \big(
	M_{-1}^{1 2}+b^{(1)}_{2} \tilde{J}_{-1}^{1 2}+b^{(1)}_{3} M_{-1}^{1 3}M_0^{3 2}+b^{(1)}_{4} \tilde{J}_{-1}^{1 3}M_0^{3 2}
	\big) |n,n'\rangle \widetilde{|1,1\rangle}\,,
	\eea
	\bea
	|\phi^{UV}_2\rangle = \big(  M_{-1}^{1 3}+b^{(2)}_2 \tilde{J}_{-1}^{1 3} \big)|n,n'\rangle \widetilde{|1,1\rangle}\,,
	\eea
	\bea
	|\phi^{UV}_3\rangle =  \big( 
	 M_{-1}^{3 2}+b^{(3)}_{2} \tilde{J}_{-1}^{3 2}+b^{(3)}_{3} M_{-1}^{1 2}M_0^{3 1}+b^{(3)}_{4} \tilde{J}_{-1}^{1 2}M_0^{3 1}+b^{(3)}_{5} M_{-1}^{2 2}M_0^{3 2}+b^{(3)}_{6} \tilde{J}_{-1}^{2 2}M_0^{3 2}+b^{(3)}_{7} M_{-1}^{3 3}M_0^{3 2}+b^{(3)}_{8} \tilde{J}_{-1}^{3 3}M_0^{3 2}
	\nonumber \\ \nonumber
	+b^{(3)}_{9} M_{-1}^{1 2}M_0^{3 2}M_0^{2 1}+b^{(3)}_{10} \tilde{J}_{-1}^{1 2}M_0^{3 2}M_0^{2 1}+b^{(3)}_{11} M_{-1}^{2 3}M_0^{3 2}M_0^{3 2}+b^{(3)}_{12} \tilde{J}_{-1}^{2 3}M_0^{3 2}M_0^{3 2}+b^{(3)}_{13} M_{-1}^{1 3}M_0^{3 2}M_0^{3 1}
	\\
	+b^{(3)}_{14} \tilde{J}_{-1}^{1 3}M_0^{3 2}M_0^{3 1}+b^{(3)}_{15} M_{-1}^{1 3}M_0^{3 2}M_0^{3 2}M_0^{2 1}+b^{(3)}_{16} \tilde{J}_{-1}^{1 3}M_0^{3 2}M_0^{3 2}M_0^{2 1}
	\big)\big)|n,n'\rangle \widetilde{|1,1\rangle}\qquad
	\eea
		\bea
	|\phi^{UV}_8\rangle = \big(
	 M_{-1}^{2 3}+b^{(8)}_{2} \tilde{J}_{-1}^{2 3}+b^{(8)}_{3} M_{-1}^{1 3}M_0^{2 1}+b^{(8)}_{4}\tilde{J}_{-1}^{13}M_0^{2 1}
	 \big)|n,n'\rangle \widetilde{|1,1\rangle}
	\eea
	\bea
	\nonumber
	|\phi^{UV}_9\rangle = \big( 
M_{-1}^{3 1}+b^{(9)}_{2} \tilde{J}_{-1}^{3 1}+b^{(9)}_{3} M_{-1}^{3 3}M_0^{3 1}+b^{(9)}_{4} \tilde{J}_{-1}^{3 3}M_0^{3 1}+b^{(9)}_{5} M_{-1}^{1 1}M_0^{3 1}+b^{(9)}_{6} \tilde{J}_{-1}^{1 1}M_0^{3 1}+b^{(9)}_{7} M_{-1}^{3 2}M_0^{2 1}+
\\\nonumber
b^{(9)}_{8} \tilde{J}_{-1}^{3 2}M_0^{2 1}+b^{(9)}_{9} M_{-1}^{2 1}M_0^{3 2}+b^{(9)}_{10} \tilde{J}_{-1}^{2 1}M_0^{3 2}+b^{(9)}_{11} M_{-1}^{1 1}M_0^{3 2}M_0^{2 1}+b^{(9)}_{12} \tilde{J}_{-1}^{1 1}M_0^{3 2}M_0^{2 1}+
\\ \nonumber
b^{(9)}_{13} M_{-1}^{3 3}M_0^{3 2}M_0^{2 1}+b^{(9)}_{14} \tilde{J}_{-1}^{3 3}M_0^{3 2}M_0^{2 1}+b^{(9)}_{15} M_{-1}^{1 2}M_0^{3 1}M_0^{2 1}+b^{(9)}_{16} \tilde{J}_{-1}^{1 2}M_0^{3 1}M_0^{2 1}+b^{(9)}_{17} M_{-1}^{2 3}M_0^{3 2}M_0^{3 1}
\\ \nonumber
+b^{(9)}_{18} \tilde{J}_{-1}^{2 3}M_0^{3 2}M_0^{3 1}+b^{(9)}_{19} M_{-1}^{1 3}M_0^{3 1}M_0^{3 1}+b^{(9)}_{20} \tilde{J}_{-1}^{1 3}M_0^{3 1}M_0^{3 1}+b^{(9)}_{21} M_{-1}^{1 2}M_0^{3 2}M_0^{2 1}M_0^{2 1}
\\\nonumber
+b^{(9)}_{22} \tilde{J}_{-1}^{1 2}M_0^{3 2}M_0^{2 1}M_0^{2 1}+b^{(9)}_{23} M_{-1}^{1 3}M_0^{3 2}M_0^{3 1}M_0^{2 1}+b^{(9)}_{24} \tilde{J}_{-1}^{1 3}M_0^{3 2}M_0^{3 1}M_0^{2 1}+b^{(9)}_{25} M_{-1}^{2 3}M_0^{3 2}M_0^{3 2}M_0^{2 1}
\\
+b^{(9)}_{26} \tilde{J}_{-1}^{2 3}M_0^{3 2}M_0^{3 2}M_0^{2 1}+b^{(9)}_{27} M_{-1}^{1 3}M_0^{3 2}M_0^{3 2}M_0^{2 1}M_0^{2 1}+b^{(9)}_{28} \tilde{J}_{-1}^{1 3}M_0^{3 2}M_0^{3 2}M_0^{2 1}M_0^{2 1}
\big)|n,n'\rangle \widetilde{|1,1\rangle} \quad
	\eea
	\bea
	\label{phiUV10}
	|\phi^{UV}_{10}\rangle = \big(
	 M_{-1}^{2 1}+b^{(10)}_{2} \tilde{J}_{-1}^{2 1}+b^{(10)}_{3} M_{-1}^{1 1}M_0^{2 1}+b^{(10)}_{4} \tilde{J}_{-1}^{1 1}M_0^{2 1}+b^{(10)}_{5} M_{-1}^{2 2}M_0^{2 1}+b^{(10)}_{6} \tilde{J}_{-1}^{2 2}M_0^{2 1}+b^{(10)}_{7} M_{-1}^{2 3}M_0^{3 1}
	 \\ \nonumber
	 +b^{(10)}_{8} \tilde{J}_{-1}^{2 3}M_0^{3 1}+b^{(10)}_{9} M_{-1}^{1 2}M_0^{2 1}M_0^{2 1}+b^{(10)}_{10} \tilde{J}_{-1}^{1 2}M_0^{2 1}M_0^{2 1}+b^{(10)}_{11} M_{-1}^{1 3}M_0^{3 1}M_0^{2 1}+b^{(10)}_{12} \tilde{J}_{-1}^{1 3}M_0^{3 1}M_0^{2 1}+
	 \\ \nonumber
	 b^{(10)}_{13} M_{-1}^{2 3}M_0^{3 2}M_0^{2 1}+b^{(10)}_{14} \tilde{J}_{-1}^{2 3}M_0^{3 2}M_0^{2 1}+b^{(10)}_{15} M_{-1}^{1 3}M_0^{3 2}M_0^{2 1}M_0^{2 1}+b^{(10)}_{16} \tilde{J}_{-1}^{1 3}M_0^{3 2}M_0^{2 1}M_0^{2 1}
	 \big)|n,n'\rangle \widetilde{|1,1\rangle}
	\eea
\end{scriptsize}
Here are their coefficients
\begin{scriptsize}
\bea
&b^{(1)}_{2}=  n-k-2,
\quad
b^{(1)}_{3}= \frac{1}{n'-1},
\quad
b^{(1)}_{4}= \frac{n -k-2}{n'-1},
\eea
\bea
b^{(2)}_2=  n+n'-k-2,
\eea
\bea
&b^{(3)}_{2}= - (k+n'+2), \quad b^{(3)}_{3}= -\frac{ n'}{(n'-1) (n+n'-1)},  \quad  b^{(3)}_{4}= \frac{ n' (k+n'+2)}{(n'-1) (n+n'-1)}  \quad 
b^{(3)}_{5}= -\frac{1}{n'-1},
\\ \nonumber 
&b^{(3)}_{6}= \frac{ k+n'+2}{n'-1},   \quad  b^{(3)}_{7}= \frac{1}{n'-1}, \;b^{(3)}_{8}= -\frac{ k+n'+2}{n'-1},  \quad  
b^{(3)}_{9}= \frac{1}{(n'-1) (n+n'-1)},  
\\ \nonumber
& b^{(3)}_{10}= -\frac{ k+n'+2}{(n'-1) (n+n'-1)},  \quad b^{(3)}_{11}= -\frac{1}{n'^2-3 n'+2},  \quad b^{(3)}_{12}= \frac{ k+n'+2}{(n'-2) (n'-1)},  \quad  b^{(3)}_{13}= -\frac{ n'}{\left(n'^2-3 n'+2\right) (n+n'-1)},  
\\ \nonumber
&b^{(3)}_{14}= \frac{n' (k+n'+2)}{(n'-2) (n'-1) (n+n'-1)}, \quad b^{(3)}_{15}= \frac{1}{\left(n'^2-3 n'+2\right) (n+n'-1)},  \quad  b^{(3)}_{16}= -\frac{ k+n'+2}{(n'-2) (n'-1) (n+n'-1)},
\eea
\bea
&b^{(8)}_{2}= n' -k-2 ,\quad b^{(8)}_{3}= -\frac{1}{n-1},\quad b^{(8)}_{4}= \frac{ k-n'+2}{n-1},
\eea
\bea
&b^{(9)}_{2}= - (k+n+n'+2), \quad b^{(9)}_{3}= \frac{ n'-2}{(n'-1) (n+n'-1)}, \quad b^{(9)}_{4}= -\frac{ (n'-2) (k+n+n'+2)}{(n'-1) (n+n'-1)}, \quad b^{(9)}_{5}= \frac{- (n'+1)}{(n'-1) (n+n'-1)},
\\ \nonumber
&b^{(9)}_{6}= \frac{ (n'+1) (k+n+n'+2)}{(n'-1) (n+n'-1)}, \quad b^{(9)}_{7}= \frac{1}{n-1}, \quad b^{(9)}_{8}= -\frac{ k+n+n'+2}{n-1}, \quad b^{(9)}_{9}= -\frac{1}{n'-1}, \quad b^{(9)}_{10}= \frac{ (k+n+n'+2)}{n'-1},
\\ \nonumber
&b^{(9)}_{11}= \frac{ 2 n+n'-3}{(n-1) (n'-1) (n+n'-1)}, \quad b^{(9)}_{12}= -\frac{ (2 n+n'-3) (k+n+n'+2)}{(n-1) (n'-1) (n+n'-1)}, \quad b^{(9)}_{13}= \frac{ n+2 n'-3}{(n-1) (n'-1) (n+n'-1)}, 
\\ \nonumber
&b^{(9)}_{14}= -\frac{ (n+2 n'-3) (k+n+n'+2)}{(n-1) (n'-1) (n+n'-1)}, \quad b^{(9)}_{15}= -\frac{ n'+1}{(n-1) (n'-1) (n+n'-1)}, \quad b^{(9)}_{16}= \frac{ (n'+1) (k+n+n'+2)}{(n-1) (n'-1) (n+n'-1)},
\\ \nonumber
&b^{(9)}_{17}= -\frac{1}{(n'-1) (n+n'-1)}, \quad b^{(9)}_{18}= \frac{ k+n+n'+2}{(n'-1) (n+n'-1)}, \quad b^{(9)}_{19}= -\frac{ n'}{(n'-1) (n+n'-2) (n+n'-1)},
 \\ \nonumber
&b^{(9)}_{20}= \frac{ n' (k+n+n'+2)}{(n'-1) (n+n'-2) (n+n'-1)}, \quad b^{(9)}_{21}= \frac{1}{(n-1) (n'-1) (n+n'-1)}, \quad b^{(9)}_{22}= -\frac{ k+n+n'+2}{(n-1) (n'-1) (n+n'-1)},
\\ \nonumber
&b^{(9)}_{23}= \frac{ n-n'-2}{(n-1) (n'-1) (n+n'-2) (n+n'-1)}, \quad b^{(9)}_{24}= \frac{- (n-n'-2) (k+n+n'+2)}{(n-1) (n'-1) (n+n'-2) (n+n'-1)}, \quad b^{(9)}_{25}= \frac{-1}{(n-1) (n'-1) (n+n'-1)}, 
\\ \nonumber
&b^{(9)}_{26}= \frac{ k+n+n'+2}{(n-1) (n'-1) (n+n'-1)}, \quad b^{(9)}_{27}= \frac{1}{(n-1) (n'-1) (n+n'-2) (n+n'-1)}, \quad b^{(9)}_{28}= \frac{- (k+n+n'+2)}{(n-1) (n'-1) (n+n'-2) (n+n'-1)},
\eea
\bea
& b^{(10)}_{2}= - (k+n+2),\quad b^{(10)}_{3}= -\frac{1}{n-1},\quad b^{(10)}_{4}= \frac{ k+n+2}{n-1},\quad b^{(10)}_{5}= \frac{1}{n-1},\quad b^{(10)}_{6}= -\frac{ (k+n+2)}{n-1},
\\ \nonumber
& b^{(10)}_{7}= \frac{1}{n+n'-1},\quad b^{(10)}_{8}= -\frac{k+n+2}{n+n'-1},\quad b^{(10)}_{9}= -\frac{1}{n^2-3 n+2},\quad b^{(10)}_{10}= \frac{k+n+2}{(n-2) (n-1)},
 \\ \nonumber
& b^{(10)}_{11}= -\frac{1}{(n-1) (n+n'-1)}, \quad  b^{(10)}_{12}= \frac{k+n+2}{(n-1) (n+n'-1)},\quad b^{(10)}_{13}= \frac{1}{(n-1) (n+n'-1)},
\\ \nonumber
 & b^{(10)}_{14}= -\frac{k+n+2}{(n-1) (n+n'-1)}, \quad b^{(10)}_{15}= -\frac{1}{(n-2) (n-1) (n+n'-1)},\quad b^{(10)}_{16}= \frac{k+n+2}{(n-2) (n-1) (n+n'-1)}\,.
\eea
\end{scriptsize}
 It is straightforward to build the $|ij\rangle$ states out of $|\phi^{IR}_{i}\rangle$ and $|\phi^{UV}_{j}\rangle$: one picks up the 
 expression for $|\phi^{UV}_{j}\rangle$ from (\ref{phiUV1})-(\ref{phiUV10})
and replaces  $|n, n'\rangle$ by the expression for  $|\phi^{IR}_{i}\rangle$ from (\ref{phiIR1})-(\ref{phiIR10}).
\section{Presentation of  the descendant fields in  coset theory}
The treatment  of descendant fields is similar.
First we notice that
\bea
\Delta ^{(k+2)}
\begin{psmallmatrix}
	n & n\\
	n' & n'	
\end{psmallmatrix}= 	
h_{k-1}(n,n')+h_{1}(1,1)
-h_{k}(n,n')
\eea
and 
\bea
\lambda_{n,n'}=\lambda_{n,n'}+\lambda_{1,1}\,.
\eea
We can express $L_{-1}$ in terms of the current operators $E$ and $J$ 
 using the Sugawara construction
\begin{small}
	\bea
	\label{Lm1phi}
	L^{IR}_{-1} |\lambda_{n,n'}\rangle|\lambda_{1,1}\rangle=
	\sum _{i,j=1}^3
	\left(
	\frac{E^{ij}_{-1}E^{ji}_0}{(k+2) (k+3)}-\frac{E^{ij}_0J^{ji}_{-1}}{k+3}
	\right)
	|\lambda_{n,n'}\rangle|\lambda_{1,1}\rangle\,.
	\eea
\end{small}
We have found also the analogues  expression for the action by  $W_{-1}$:
\begin{scriptsize}
	\bea
	\label{Wm1phi}
	W^{IR}_{-1} |\lambda_{n,n'}\rangle|\lambda_{1,1}\rangle=\frac{1}{\sqrt{6 (2 k+1) (2 k+9)} (k+3)}\bigg[
	\sum_{i,j=1}^{3}\left(\frac{9 E^{ij}_{-1}E^{ji}_0}{(k+2)}-9 E_0^{ij}J_{-1}^{ji}\right)
	\\ \nonumber
	+
	\sum_{i,j,k=1}^{3}	\left(
	+6J^{ij}_{-1}E^{jk}_0E^{ki}_0-\frac{6}{(k+2)}E^{ij}_{-1}E^{jk}_0E^{ki}_0
	\right)\bigg]
	|\lambda_{n,n'}\rangle|\lambda_{1,1}\rangle \,.
	\eea
\end{scriptsize}
The  descendant fields of our interest  are appropriate combinations of (\ref{Lm1phi}) and (\ref{Wm1phi}).
\section{Explicit expressions for ${\cal M}$ }
\label{M}
Expressions for some states (e.g.  $| 29 \rangle$) are very large and the computation of inner products 
is very cumbersome. Fortunately there is a procedure that substantially  improving the situation. Since
\bea
\label{an_bra}
\langle ij | \left(E^{kl}_n+J^{kl}_n+\tilde{J}^{kl}_n\right)=0\,
\quad
{\rm for}
\quad
n<0\,,
\\
\langle ij | \left(E^{kl}_0+J^{kl}_0+\tilde{J}^{kl}_0\right)=0\,
\quad
{\rm for}
\quad
k>l\,,
\eea
in $| ij \rangle$ we can subsequently move all the operators $E^{kl}_n$ to the leftmost position and replace them by  
$-(J^{kl}_n +\tilde{J}^{kl}_n)$. Now we can ignore all terms
in  $\langle ij |$ including operators $E$, since they are free to move to the left and eventually kill the primary bra state \footnote{For descendant states
$| ij \rangle$ with $i,j=4,5,6,7$ one should replace operators $E^{kk}_0$  by their  eigenvalues in advance.}.
Here are our results
\begin{scriptsize}
	\bea
	&{\cal M}_{11}=\frac{1}{n+n'+1},
	\quad
	{\cal M}_{21}=
	\frac{k^2+k (n'+6)-2 \left(n^2+(n-1) n'+n-4\right)}{(n+1) (k-n+2) (n+n'+1) (k+n+n'+4)},
	\quad
	{\cal M}_{31}=1,
	\eea
	\bea
	 {\cal M}_{41}=\frac{3 k^2 (n-2)-3 k (n-5) (n-2)-2 \left(4 n^2+n (n'-15)+n'^2+15\right)}{2 (k+3) (k-n+2) \left(n^2+n n'+n'^2-3\right)},
	\eea
	\bea
	&{\cal M}_{51}=-\frac{(k+2) (n-1) (k-n+3) (n+2 n')}{2 (k+3) (k-n+2) \left(n^2+n n'+n'^2-3\right)}
	\sqrt{\frac{(6 k+3) (2 k+9) (n'+1) (n+n'-1)}{(2 k+1) (2 k+9) \left(n^2-1\right) (n'-1) (n+n'+1)}},
	\eea
	\bea
	&{\cal M}_{71}=
	\frac{1}{2 (k+3) (k-n+2)}
	\left(
	\frac{2 (k+2) (k-n+3)}{(n+1) (n+n'+1)}
	-\frac{3 (k+2) (n-2) (k-n+3)}{n^2+n n'+n'^2-3}-\frac{2 (k+2) (k-n+3)}{(n+1) (n'-1)}+\frac{4 k (k-n+5)-10 n+22}{n+1}
	\right),
	\eea
	\bea
     &	{\cal M}_{81}=
	\frac{-(n+2) \left(k (k+6)-2 \left(n^2+n-4\right)\right)+n'^2 (2 k (k+6)-3 n (n+1)+16)+n n' (2 k (k+6)-3 n (n+1)+16)}{(n+1) (n+2) (n'-1) (k-n+2) (k+n+4) (n+n'+1)},
	\eea
	\bea
		& {\cal M}_{91}=\frac{1}{1-n'},
		\quad
		{\cal M}_{101}=\frac{n' (k-2 n+2)-(k+4) (k-n+2)}{(n+1) (n'-1) (k-n+2) (k-n'+4)},
		\quad
		{\cal M}_{12}=\frac{(k+4) (k-n+2)-2 (n+1) n'-2 n'^2}{(n'+1) (k+n'+4) (n+n'+1) (k-n-n'+2)},
	\eea
	\bea
	& {\cal M}_{22}=
	\frac{-n \left(2 (k (k+6)+2) n'+k (k+6)-3 n'^3-9 n'^2+4\right)+(n'+2) (2 n' (n'+1)-(k+2) (k+4))+n^3 (3 n'+2)+n^2 \left(6 n'^2+9 n'+6\right)}{(n+1) (n'+1) (n+n'+1) (n+n'+2) (-k+n+n'-2) (k+n+n'+4)},
	\eea
	\bea
	&{\cal M}_{32}=\frac{1}{n'+1},
	\quad
	{\cal M}_{42}=
	\frac{3 k^2 (n+n'-2)-3 k (n+n'-5) (n+n'-2)-2 \left(4 n^2+7 n n'+4 n'^2\right)+30 (n+n'-1)}{2 (k+3) \left(n^2+n n'+n'^2-3\right) (k-n-n'+2)},
	\eea
	\bea
	& {\cal M}_{52}=
	-\frac{ (k+2) (n-1) (n'-1) (n-n') (-k+n+n'-3)}{2 (k+3) \left(n^2+n n'+n'^2-3\right) (k-n-n'+2)}
	\sqrt{\frac{3(n+n'-1)}{\left(n^2-1\right) \left(n'^2-1\right) (n+n'+1)}},
	\eea
	\bea
	 {\cal M}_{72}=\quad\left(\quad 2 (k+3) \left(n+1\right) \left(n'+1\right) (n+n'+1) \left(n^2+n n'+n'^2-3\right) (-k+n+n'-2)\quad \right)^{-1}
	  \quad
	 \times
	 \\ \nonumber
	 \bigg[\quad
	 -k^2 \left(n^3 (n'+3)-2 n^2 (n'-4) (n'+1)+n (n'-1) (n' (n'+7)+9)+(n'+3) (n' (3 n'-1)-6)\right)\quad+
	 \\ \nonumber
	 k (n+n'-5) \left(n^3 (n'+3)-2 n^2 (n'-4) (n'+1)+n (n'-1) (n' (n'+7)+9)+(n'+3) (n' (3 n'-1)-6)\right)
	 \\ \nonumber
	 +4 n^4 n'+20 n^3 n'-14 n^2 n'+8 n^4+2 n^3-70 n^2-14 n n'^2-64 n n'+6 n-70 n'^2+6 n'+102
	 \\ \nonumber
	 +2 n^2 n'^3+2 n^3 n'^2+46 n^2 n'^2+4 n n'^4+20 n n'^3+8 n'^4+2 n'^3\quad
	 \bigg]\,,
	\eea
	\bea
{\cal M}_{82}=
\frac{(k+4) (-k+n'-2)+2 n^2+2 n (n'+1)}{(n+1) (k+n+4) (n+n'+1) (-k+n+n'-2)},
\quad
{\cal M}_{102}=
\frac{1}{n+1},
\eea
\bea
&
{\cal M}_{33}=-\frac{1}{n+n'-1},
\quad
{\cal M}_{43}=
-\frac{(3 k+8) n'^2+3 (k+2) (k+5) n'+6 k (k+5)+2 n^2+2 n n'+30}{2 (k+3) (k+n'+2) \left(n^2+n n'+n'^2-3\right)},
\eea
\bea
&{\cal M}_{53}=-
\frac{((k+2) (n-1) (n'+1) (k+n'+3) (2 n+n')) }{(k+3) (k+n'+2) \left(n^2+n n'+n'^2-3\right)}
\sqrt{\frac{3 (2 k+9) (2 k+1) (n+n'+1)}{2 \left(8 k^2+40 k+18\right) \left(n^2-1\right) \left(n'^2-1\right) (n+n'-1)}},
\eea
\bea
{\cal M}_{73}=-\left(\quad 2 (k+3) \left(n+1\right) \left(n'+1\right) (k+n'+2) (n+n'-1) \left(n^2+n n'+n'^2-3\right)\quad \right)^{-1} \quad
\times
\\ \nonumber
\bigg[\,
k^2 \left(8 n^3 n'+n^2 (n'+2) (5 n'-7)+4 n^4+n n' (n' (n'+3)-14)+(n'-3) \left(3 n'^2+n'-6\right)\right)\quad+
\\ \nonumber
k (n'+5) \left(8 n^3 n'+n^2 (n'+2) (5 n'-7)+4 n^4+n n' (n' (n'+3)-14)+(n'-3) \left(3 n'^2+n'-6\right)\right)
\\ \nonumber
+34 n^2 n'^2+10 n^4 n'+44 n^3 n'-20 n^2 n'+22 n^4-76 n^2-20 n n'^2-76 n n'-70 n'^2-6 n'+102
\\ \nonumber
+14 n^2 n'^3++20 n^3 n'^2+4 n n'^4+12 n n'^3+8 n'^4-2 n'^3
\quad \bigg]\,,
\eea
\bea
& {\cal M}_{83}=
-\frac{n (k+2 n'+2)+(k+4) (k+n'+2)}{(n+1) (n'-1) (k+n+4) (k+n'+2)},
\quad
{\cal M}_{93}=
\frac{(k+2) (k-n+4)-2 (n-1) n'-2 n'^2}{(n'-1) (k+n'+2) (n+n'-1) (k-n-n'+4)},
\eea
\bea
&{\cal M}_{103}=
\frac{n' \left(k^2+2 (k+2) (k+4) n+6 k+3 n^2+4\right)+2 (k+2) (k+4) \left(n^2-1\right)-\left((3 n+2) n'^3\right)-3 (n-2) (n+1) n'^2}{(n+1) (n'-2) (n'-1) (k-n'+4) (k+n'+2) (n+n'-1)},
\eea
\bea
&{\cal M}_{14}=
\frac{3 k^2 (n'-2)+3 k (n'-2) (n'+7)-2 \left(n^2+n n'-n' (5 n'+6)+33\right)}{2 (k+3) (k+n'+4) \left(n^2+n n'+n'^2-3\right)},
\eea
\bea
&{\cal M}_{24}=
\frac{3 k^2 (n+n'-2)+3 k (n+n'-2) (n+n'+7)+2 \left(5 n^2+n (11 n'+6)+n' (5 n'+6)-33\right)}{2 (k+3) \left(n^2+n n'+n'^2-3\right) (k+n+n'+4)},
\eea
\bea
&{\cal M}_{34}=
-\frac{3 k^2 (n+2)-3 k (n-7) (n+2)+2 \left(-5 n^2+n (n'+6)+n'^2+33\right)}{2 (k+3) (k-n+4) \left(n^2+n n'+n'^2-3\right)},
\eea
\bea
&{\cal M}_{44}=
\frac{k (k+6) \left(2 n^2-15\right)+2 (k (k+6)+7) n n'+2 (k (k+6)+7) n'^2+2 \left(7 n^2-57\right)}{2 (k+3)^2 \left(n^2+n n'+n'^2-3\right)},
\quad
{\cal M}_{54}=0,
\eea
\bea
& {\cal M}_{74}=
-\frac{9 k (k+6)+2 \left(n^2+n n'+n'^2+33\right)}{2 (k+3)^2 \left(n^2+n n'+n'^2-3\right)},
\quad
{\cal M}_{84}=
\frac{3 k^2 (n-2)+3 k (n-2) (n+7)-2 n n'+2 n (5 n+6)-2 n'^2-66}{2 (k+3) (k+n+4) \left(n^2+n n'+n'^2-3\right)},
\eea
\bea
& {\cal M}_{94}=
-\frac{3 k^2 (n+n'+2)-3 k (n+n'-7) (n+n'+2)-2 \left(5 n^2+n (11 n'-6)+n' (5 n'-6)-33\right)}{2 (k+3) \left(n^2+n n'+n'^2-3\right) (k-n-n'+4)},
\eea
\bea
& {\cal M}_{104}=
-\frac{3 k^2 (n'+2)-3 k (n'-7) (n'+2)+2 \left(n^2+n n'+n' (6-5 n')+33\right)}{2 (k+3) (k-n'+4) \left(n^2+n n'+n'^2-3\right)},
\eea
\bea
& {\cal M}_{15}=
\frac{(k+4) (n+1) (n'-1) (k+n'+3) (2 n+n') }{(k+3) (k+n'+4) \left(n^2+n n'+n'^2-3\right)}
\sqrt{\frac{3 (2 k+3) (2 k+11) (n+n'-1)}{2 (8 k (k+7)+66) \left(n^2-1\right) \left(n'^2-1\right) (n+n'+1)}},
\eea
\bea
& {\cal M}_{25}=
 \frac{(k+4) (n-n') (k+n+n'+3) }{(k+3) \left(n^2+n n'+n'^2-3\right) (k+n+n'+4)}
 \sqrt{\frac{3 (2 k+3) (2 k+11) (n-1) (n'-1) (n+n'-1)}{2 (8 k (k+7)+66) (n+1) (n'+1) (n+n'+1)}},
\eea
\bea
&{\cal M}_{35}=
\frac{(k+4) (n+1) (k-n+3) (n+2 n') }{(k+3) (k-n+4) \left(n^2+n n'+n'^2-3\right)}
\sqrt{\frac{3 (2 k+11) (2 k+3) (n'-1) (n+n'+1)}{2 (8 k (k+7)+66) \left(n^2-1\right) (n'+1) (n+n'-1)}},
\eea
\bea
&{\cal M}_{55}=
\frac{(k+2) (k+4) \left(2 \left(n^2+n n'+n'^2\right)-15\right)}{2 (k+3)^2 \left(n^2+n n'+n'^2-3\right)},
\quad
{\cal M}_{65}=-\frac{9 (k+2) (k+4)}{2 (k+3)^2 \left(n^2+n n'+n'^2-3\right)},
\\
&{\cal M}_{75}=
\frac{\sqrt{3} (k+2) (k+4) (n'-n) (2 n+n') (n+2 n')}{(k+3)^2 \left(n^2+n n'+n'^2-3\right) \sqrt{\left(n^2-1\right) \left(n'^2-1\right) \left((n+n')^2-1\right)}},
\eea
\bea
&{\cal M}_{85}=
\frac{(k+4) (1-n) (n'+1) (k+n+3) (n+2 n') }{(k+3) (k+n+4) \left(n^2+n n'+n'^2-3\right)}
\sqrt{\frac{3 (2 k+11) (2 k+3) (n+n'-1)}{2 (8 k (k+7)+66) \left(n^2-1\right) \left(n'^2-1\right) (n+n'+1)}},
\eea
\bea
&{\cal M}_{95}=
\frac{(k+4) (n+1) (n'+1) (n'-n) (-k+n+n'-3) }{(k+3) \left(n^2+n n'+n'^2-3\right) (-k+n+n'-4)}
\sqrt{\frac{3 (2 k+11) (2 k+3) (n+n'+1)}{2 (8 k (k+7)+66) \left(n^2-1\right) \left(n'^2-1\right) (n+n'-1)}},
\eea
\bea
&{\cal M}_{105}=
-\frac{(k+4) (n'+1) (k-n'+3) (2 n+n') }{(k+3) (k-n'+4) \left(n^2+n n'+n'^2-3\right)}
\sqrt{\frac{3 (2 k+3) (2 k+11) (n-1) (n+n'+1)}{2 (8 k (k+7)+66) (n+1) \left(n'^2-1\right) (n+n'-1)}},
\eea
\bea
{\cal M}_{17}=\quad \left(\, 2 (k+3) \left(n+1\right) \left(n'+1\right) (k+n'+4) (n+n'+1) \left(n^2+n n'+n'^2-3\right)\, \right)^{-1}
\quad
\times
\\ \nonumber
\bigg[ \;
k^2 \left(2 \left(2 n^4-7 n^2+9\right)+(n-3) n'^3+(n+1) (5 n-8) n'^2+(n-1) (n (8 n+5)-9) n'\right)\quad+
\\ \nonumber
k (n'+7) \left(2 \left(2 n^4-7 n^2+9\right)+(n-3) n'^3+(n+1) (5 n-8) n'^2+(n-1) (n (8 n+5)-9) n'\right)
\\ \nonumber
+14 n^4 n'+92 n^3 n'-82 n^2 n'+46 n^4-160 n^2-82 n n'^2-160 n n'-64 n'^2+168 n'+210
\\ \nonumber
+16 n^2 n'^3+28 n^3 n'^2++46 n^2 n'^2+2 n n'^4-10 n'^4-64 n'^3
\quad
\bigg]\,,
\eea
\bea
{\cal M}_{27}\quad = \left(\, 2 \, (k+3) \, (n+1) \, (n'+1) \, (n+n'+1) \, \left(n^2+n n'+n'^2-3\right)\,  (k+n+n'+4)\, \right)^{-1} \quad \times
\\ \nonumber
\bigg[ \quad
k^2 \, \left(\, n^3 (n'+3)-2 n^2 (n'-4) (n'+1)+n (n'-1) (n' (n'+7)+9)+(n'+3) (n' (3 n'-1)-6)\, \right)\,+
\\ \nonumber
k (n+n'+7) \left(n^3 (n'+3)-2 n^2 (n'-4) (n'+1)+n (n'-1) (n' (n'+7)+9)+(n'+3) (n' (3 n'-1)-6)\right)
\\ \nonumber
+2 n^4 n'+40 n^3 n'+110 n^2 n'+10 n^4+64 n^3+64 n^2-32 n n'-168 n+64 n'^2-168 n'-210
\\ \nonumber
-8 n^2 n'^3-8 n^3 n'^2+14 n^2 n'^2+2 n n'^4+40 n n'^3++110 n n'^2+10 n'^4+64 n'^3
\bigg]\,,
\eea
\bea
{\cal M}_{37}=\frac{1}{2 (k+3) (k-n+4)}
\bigg[14+4 k-\frac{4 (k+2) (k+4)}{n-1}+
\frac{3 (k+4) (n+2) (k-n+3)}{n^2+n n'+n'^2-3}
\\ \nonumber
-\frac{2 (k+4) (k-n+3)}{(n-1) (n'+1)}+\frac{2 (k+4) (k-n+3)}{(n-1) (n+n'-1)}\bigg]\,,
\eea
\bea
&{\cal M}_{77}=\frac{1}{2 (k+3)^2}\bigg[
\frac{2 \left(k (k+6) \left(n^2-5\right)+7 n^2-39\right)}{n^2-1}+\frac{2 (k+2) (k+4) (2 (n-1) n-1)}{(n-1) n (n'+1)}-\frac{2 (k+2) (k+4) (2 n (n+1)-1)}{n (n+1) (n'-1)}
\\ \nonumber
&\qquad \qquad \qquad \qquad\qquad+\frac{27 (k+2) (k+4)}{n^2+n n'+n'^2-3}-\frac{2 (k+2) (k+4) (2 (n-1) n-1)}{(n-1) n (n+n'-1)}+\frac{2 (k+2) (k+4) (2 n (n+1)-1)}{n (n+1) (n+n'+1)}
\bigg]\,,
\eea
\bea
&{\cal M}_{87}=\frac{1}{2 (k+3) (k+n+4)}\bigg[\frac{2 (k+4) (k+n+3)}{(n+1) (n+n'+1)}+\frac{4 k (k+n+7)+14 n+46}{n+1}-\frac{3 (k+4) (n-2) (k+n+3)}{n^2+n n'+n'^2-3}-\frac{2 (k+4) (k+n+3)}{(n+1) (n'-1)}\bigg],
\eea
\bea
{\cal M}_{97}\,=\,\left(\,2 \,(k+3)\, (n-1)\, (n'-1)\, (n+n'-1)\, \left(n^2+n n'+n'^2-3\right)\, (k-n-n'+4)\,\right)^{-1}\,\,
\times \\
\bigg[
k^2 \left(2 n^2 (n'-1) (n'+4)-n^3 (n'-3)-n (n'+1) ((n'-7) n'+9)+(n'-3) \left(3 n'^2+n'-6\right)\right)
\\ \nonumber
k (n+n'-7) \left(n^3 (n'-3)-2 n^2 (n'-1) (n'+4)+n (n'+1) ((n'-7) n'+9)-(n'-3) \left(3 n'^2+n'-6\right)\right)
\\ \nonumber
+2 n^4 n'-40 n^3 n'+110 n^2 n'-10 n^4+64 n^3-64 n^2+32 n n'-168 n-64 n'^2-168 n'+210
\\ \nonumber
-8 n^2 n'^3-8 n^3 n'^2-14 n^2 n'^2+2 n n'^4-40 n n'^3+110 n n'^2-10 n'^4+64 n'^3
\bigg],
\eea
\bea
{\cal M}_{107}\,=\, \left(\, 2 \, (k+3)\,  (n+1) \, (n'-1) \, (k-n'+4)\,  (n+n'-1)\,  \left(n^2+n n'+n'^2-3\right)\, \right)^{-1}\, 
\times
\\ \nonumber
\bigg[
-k^2 \left(8 n^3 n'+n^2 (n'+2) (5 n'-7)+4 n^4+n n' (n' (n'+3)-14)+(n'-3) \left(3 n'^2+n'-6\right)\right)+
\\ \nonumber
k (n'-7) \left(8 n^3 n'+n^2 (n'+2) (5 n'-7)+4 n^4+n n' (n' (n'+3)-14)+(n'-3) \left(3 n'^2+n'-6\right)\right)
\\ \nonumber
+14 n^4 n'-92 n^3 n'-82 n^2 n'-46 n^4+160 n^2+160 n n'+64 n'^2+168 n'-210
\\ \nonumber
+16 n^2 n'^3+28 n^3 n'^2-46 n^2 n'^2+2 n n'^4-82 n n'^2+10 n'^4-64 n'^3
\bigg],
\eea
\bea
&{\cal M}_{18}=
\frac{\left(n'^2-(k+3)^2\right) \left(n' \left(k^2-2 (k+2) (k+4) n+6 k+3 n^2+4\right)-2 (k+2) (k+4) \left(n^2-1\right)+(3 n-2) n'^3+3 (n-1) (n+2) n'^2\right)}{(n-1) (n'+1) (n'+2) (k-n'+2) (k-n'+3) (k+n'+3) (k+n'+4) (n+n'+1)},
\eea
\bea
&{\cal M}_{28}=
\frac{(k+2) (k+n+4)-2 (n+1) n'-2 n'^2}{(n'+1) (k-n'+2) (n+n'+1) (k+n+n'+4)},
\quad
{\cal M}_{38}=
-\frac{(k+2) (k-n+4)-n' (k-2 n+4)}{(n-1) (n'+1) (k-n+4) (k-n'+2)},
\eea
\bea
{\cal M}_{48}=
-\frac{-n' (3 k (k+7)-2 n+30)+(3 k+8) n'^2+6 k (k+5)+2 n^2+30}{2 (k+3) (k-n'+2) \left(n^2+n n'+n'^2-3\right)},
\eea
\bea
&{\cal M}_{58}=
\frac{((k+2) (n'-1) (k-n'+3) (2 n+n')) }{(k+3) (k-n'+2) \left(n^2+n n'+n'^2-3\right)}
\sqrt{\frac{3 (2 k+9) (2 k+1) (n+1) (n+n'-1)}{2 \left(8 k^2+40 k+18\right) (n-1) \left(n'^2-1\right) (n+n'+1)}},
\eea
\bea
{\cal M}_{78}\, =\,  \left(\,  2\,   (k+3)\,  (n-1) \, (n'+1) \, (k-n'+2)\,  (n+n'+1)\textbf{} \left(n^2+n n'+n'^2-3\right)\, \right)^{-1}\, 
\times
\\ \nonumber
\bigg[
k^2 \left(2 \left(2 n^4-7 n^2+9\right)+(n-3) n'^3+(n+1) (5 n-8) n'^2+(n-1) (n (8 n+5)-9) n'\right)-
\\ \nonumber
k (n'-5) \left(2 \left(2 n^4-7 n^2+9\right)+(n-3) n'^3+(n+1) (5 n-8) n'^2+(n-1) (n (8 n+5)-9) n'\right)
\\ \nonumber
-10 n^4 n'+44 n^3 n'+20 n^2 n'+22 n^4-76 n^2+20 n n'^2-76 n n'-70 n'^2+6 n'+102
\\ \nonumber
-14 n^2 n'^3-20 n^3 n'^2++34 n^2 n'^2-4 n n'^4+12 n n'^3+8 n'^4+2 n'^3
\bigg]\,,
\eea
\bea
&{\cal M}_{98}=-\frac{1}{n-1},
\quad
{\cal M}_{39}=\frac{(-k+n-3) \left(-(k+4) (k+n'+2)+2 n^2+2 n (n'-1)\right)}{(n-1) (k-n+3) (k-n+4) (n+n'-1) (k+n+n'+2)},
\eea
\bea
&{\cal M}_{49}=-\frac{3 k^2 (n+n'+2)+3 k (n+n'+2) (n+n'+5)+8 n^2+14 n n'+30 (n+n'+1)+8 n'^2}{2 (k+3) \left(n^2+n n'+n'^2-3\right) (k+n+n'+2)},
\eea
\bea
&{\cal M}_{59}=
\frac{(k+2) (n'-n) (k+n+n'+3) }{(k+3) \left(n^2+n n'+n'^2-3\right) (k+n+n'+2)}
\sqrt{\frac{3 (2 k+9) (2 k+1) (n+1) (n'+1) (n+n'+1)}{2 \left(8 k^2+40 k+18\right) (n-1) (n'-1) (n+n'-1)}},
\eea
\bea
{\cal M}_{79}\, =\, -\, \left(\, 2\,  (k+3) \, (n-1)\,  (n'-1)\,  (n+n'-1) \, \left(n^2+n n'+n'^2-3\right)\,  (k+n+n'+2)\, \right)^{-1}\, \times
\\ \nonumber
\bigg[\, 
k^2 \left(n^3 (n'-3)-2 n^2 (n'-1) (n'+4)+n (n'+1) ((n'-7) n'+9)-(n'-3) \left(3 n'^2+n'-6\right)\right)\,+
\\ \nonumber
k (n+n'+5) \left(n^3 (n'-3)-2 n^2 (n'-1) (n'+4)+n (n'+1) ((n'-7) n'+9)-(n'-3) \left(3 n'^2+n'-6\right)\right)
\\ \nonumber
+4 n^4 n'-20 n^3 n'-14 n^2 n'-8 n^4+2 n^3+70 n^2-14 n n'^2+64 n n'+6 n+70 n'^2+6 n'-102
\\ \nonumber
+2 n^2 n'^3+2 n^3 n'^2-46 n^2 n'^2+4 n n'^4-20 n n'^3-8 n'^4+2 n'^3
\, 
\bigg],
\eea
\bea
& {\cal M}_{99}=
\frac{n' \left(k^2-2 (k (k+6)+2) n+6 k+3 n^3-9 n^2+4\right)+(n-2) (k (k+6)-2 (n-1) n+8)+(3 n-2) n'^3+3 (n (2 n-3)+2) n'^2}{(n-1) (n'-1) (n+n'-2) (n+n'-1) (-k+n+n'-4) (k+n+n'+2)},
\eea
\bea
& {\cal M}_{109}=
\frac{(k+4) (k+n+2)-2 (n-1) n'-2 n'^2}{(n'-1) (k-n'+4) (n+n'-1) (k+n+n'+2)},
\eea
\bea
& {\cal M}_{110}=-\frac{n' (k+2 n+2)+(k+4) (k+n+2)}{(n-1) (n'+1) (k+n+2) (k+n'+4)},
\eea
\bea
& {\cal M}_{310}=
\frac{n'^2 (2 k (k+6)-3 (n-1) n+16)+n n' (2 k (k+6)-3 (n-1) n+16)+(n-2) (k (k+6)-2 (n-1) n+8)}{(n-2) (n-1) (n'+1) (k-n+4) (k+n+2) (n+n'-1)},
\eea
\bea
& {\cal M}_{410}=
-\frac{3 k^2 (n+2)+3 k (n+2) (n+5)+2 \left(4 n^2+n (n'+15)+n'^2+15\right)}{2 (k+3) (k+n+2) \left(n^2+n n'+n'^2-3\right)},
\eea
\bea
&{\cal M}_{510}=
\frac{ (k+2) (n+1) (n'-1) (k+n+3) (n+2 n')}{(k+3) (k+n+2) \left(n^2+n n'+n'^2-3\right)}
\sqrt{\frac {3 (2 k+9) (2 k+1) (n+n'+1)}{2\left(8 k^2+40 k+18\right) \left(n^2-1\right) \left(n'^2-1\right) (n+n'-1)}},
\eea
\bea
& {\cal M}_{710}=\frac{1}{2 (k+3) (k+n+2)}\bigg[
\frac{3 (k+2) (n+2) (k+n+3)}{n^2+n n'+n'^2-3}-\frac{2 (k+2) (k+n+3)}{(n-1) (n'+1)}+\frac{2 (k+2) (k+n+3)}{(n-1) (n+n'-1)}-\frac{2 (2 k (k+n+5)+5 n+11)}{n-1}
\bigg],
\eea
\bea
{\cal M}_{910}=
\frac{-k^2+k (n'-6)+2 (n (n+n'-1)+n'-4)}{(n-1) (k+n+2) (n+n'-1) (-k+n+n'-4)},
\eea
\end{scriptsize}
and
\begin{scriptsize}
	\bea
	&{\cal M}_{55}={\cal M}_{66}\,,
	\quad
	{\cal M}_{56}={\cal M}_{65},
	\\ \nonumber
	&{\cal M}_{i5}={\cal M}_{i6},
	\quad
	{\cal M}_{5i}={\cal M}_{6i},
	\quad \text{for} \quad
	i=1,2,3,4,7,8,9, 10 \,,
	\eea
	\bea
	&	{\cal M}_{11}=	{\cal M}_{88}\,,
	\quad
	{\cal M}_{31}=	{\cal M}_{92}=	{\cal M}_{13}=	{\cal M}_{108}={\cal M}_{29}=	{\cal M}_{810}\,,
	\\ \nonumber
	&	{\cal M}_{91}={\cal M}_{89},
	\quad
	{\cal M}_{32}={\cal M}_{210},
	\quad
	{\cal M}_{102}={\cal M}_{23},
	\quad
	{\cal M}_{33}={\cal M}_{1010},
	\\ \nonumber
	&	{\cal M}_{54}={\cal M}_{64}={\cal M}_{45}={\cal M}_{46},
	\quad
	{\cal M}_{47}={\cal M}_{74},
	\quad
	{\cal M}_{57}={\cal M}_{67}={\cal M}_{75}={\cal M}_{76},
	\quad
	{\cal M}_{98}={\cal M}_{19}\,.
	\eea
\end{scriptsize}
\end{appendix}
\bibliographystyle{JHEP}

\providecommand{\href}[2]{#2}\begingroup\raggedright\endgroup

\end{document}